\documentclass[twocolumn]{aastex631}

\newcommand{\teff}{$T$}
\newcommand{\logg}{$\log$ g}
\newcommand{\vsini}{$v\sin i$}

\newcommand{\rk}{$r_{\rm K}$}

\newcommand{\alfa}{$\alpha_{K_s-W4}$}
\newcommand{\ext}{A$_{\rm V}$}

\defcitealias{lopezv21}{Paper I}
\defcitealias{kidder2021}{Paper II}
\submitjournal{ApJ}

\shortauthors{L\'opez-Valdivia et al.}
\shorttitle{}

\usepackage{url}
\usepackage{amsmath}

\begin{document}

 
\title{The IGRINS YSO Survey III: Stellar parameters of pre-main sequence stars in Ophiuchus and Upper Scorpius}


\author[0000-0002-7795-0018]{Ricardo L\'opez-Valdivia}
\affil{The University of Texas at Austin,
Department of Astronomy, 
2515 Speedway, Stop C1400,  
Austin, TX 78712-1205}
\email{rlopezv@astro.unam.mx}
\affil{Universidad Nacional Aut\'onoma de M\'exico, Instituto de Astronom\'ia, AP 106,  Ensenada 22800, BC, M\'exico}

\author[0000-0001-7875-6391]{Gregory N. Mace}
\affil{The University of Texas at Austin,
Department of Astronomy, 
2515 Speedway, Stop C1400,
Austin, TX 78712-1205}

\author[0000-0001-9797-0019]{Eunkyu Han}
\affil{The University of Texas at Austin,
Department of Astronomy,
2515 Speedway, Stop C1400,
Austin, TX 78712-1205}

\author[0000-0002-8378-1062]{Erica Sawczynec}
\affil{The University of Texas at Austin,
Department of Astronomy,
2515 Speedway, Stop C1400,
Austin, TX 78712-1205}

\author[0000-0001-9797-5661]{Jes\'us Hern\'andez}
\affiliation{Universidad Nacional Aut\'onoma de M\'exico, Instituto de Astronom\'ia, AP 106,  Ensenada 22800, BC, M\'exico}

\author[0000-0001-7998-226X]{L. Prato}
\affil{Lowell Observatory, 1400 W. Mars Hill Rd, Flagstaff. AZ 86001. USA}

\author{Christopher M. Johns-Krull}
\affil{Physics \& Astronomy Dept., Rice University,
6100 Main St., Houston, TX 77005}

\author[0000-0002-0418-5335]{Heeyoung Oh}
\affil{Korea Astronomy and Space Science Institute, 776 Daedeok-daero, 
Yuseong-gu, Daejeon 34055, Korea}

\author{Jae-Joon Lee}
\affil{Korea Astronomy and Space Science Institute, 776 Daedeok-daero, 
Yuseong-gu, Daejeon 34055, Korea}

\author{Adam Kraus}
\affil{The University of Texas at Austin,
Department of Astronomy,
2515 Speedway, Stop C1400,
Austin, TX  78712-1205}

\author{Joe Llama}
\affil{Lowell Observatory, 1400 W. Mars Hill Rd, Flagstaff. AZ 86001. USA}

\author[0000-0003-3577-3540]{Daniel T. Jaffe}
\affil{The University of Texas at Austin,
Department of Astronomy, 2515 Speedway, Stop C1400, Austin, TX  78712-1205}

\begin{abstract}
We used the Immersion GRating Infrared Spectrometer (IGRINS) to determine fundamental parameters for {61} K- and M-type young stellar objects (YSOs) located in the Ophiuchus and Upper Scorpius star-forming regions. We employed synthetic spectra and a Markov chain Monte Carlo approach to fit specific K-band spectral regions and determine the photospheric temperature (\teff), surface gravity (\logg), magnetic field strength (B), projected rotational velocity (\vsini), {and K-band veiling (\rk)}. We determined B for {$\sim$46\%} of our sample. Stellar parameters were compared to the results from Taurus-Auriga and the TW Hydrae Association (TWA) presented in \citetalias{lopezv21} of this series. We classified all the YSOs in the IGRINS survey with infrared spectral indices from 2MASS and WISE photometry between 2 and 24~$\mu$m. We found that Class {\sc ii} YSOs typically have lower \logg\ and \vsini, similar B, and higher K-band veiling than their Class {\sc iii} counterparts. Additionally, we determined the stellar parameters for a sample of K and M field stars also observed with IGRINS.
We have identified intrinsic similarities and differences at different evolutionary stages with our homogeneous determination of stellar parameters in the IGRINS YSO Survey. { Considering \logg\ as a proxy for age, we found that the Ophiuchus and Taurus samples have a similar age. We also find that Upper Scorpius and TWA YSOs have similar ages, and are more evolved than Ophiuchus/Taurus YSOs.}
\end{abstract}


%-----------

\section{Introduction}
{{Young stellar objects (YSOs) are stars at an early stage of evolution and their properties have a dominant impact on the environments of planet development.
Stellar parameters like temperature (\teff) and surface gravity (\logg) permit comparisons with evolutionary models \citep{baraffe15, simon19}. 
With these fundamental parameters, the differences between models and observations for stars with masses $<$1.5 M$_{\odot}$ and age $<$10 Myr (e.g., \citealp{simon13}) can be better understood. 
The accurate and precise determination of YSO parameters is limited by the impacts of interstellar reddening, continuum veiling, and magnetic fields. These processes, if not considered, can result in inaccurate YSO parameters and thus lead to wrong conclusions about star formation and evolution.}  

For example, the veiling is a non-stellar continuum emission that reduces the depth of the photospheric lines in YSO spectra \citep{joy49}. Such reduction of the line depth could lead to erroneous \teff\ or \logg\ values as these parameters have similar effects on certain atomic lines. On the other hand, the presence of strong magnetic fields (B) changes, through a Zeeman broadening,  the absorption line profiles  \citep[e.g.,][]{johns-krull99,yang05,lavail17,sokal18,lavail19,sokal20}. }

The Immersion GRating INfrared Spectrometer \citep[IGRINS;][]{yuk10,park14,mace16} was designed to minimize the problems associated with determining YSO parameters. IGRINS employs a silicon immersion grating as the primary disperser \citep[][]{jaffe98, wang10, gully10} and volume phase holographic gratings to cross-disperse the H- and K-band echellograms onto Teledyne Hawaii-2RG arrays. This setup provides a compact design with high sensitivity and a significant single-exposure spectral grasp at a spectral resolution of $R\sim 45000$ \citep{yuk10,park14}. 
IGRINS has a fixed spectral format and no moving optics, so the science products are consistent over time baselines of several years. IGRINS has increased its scientific value by traveling between McDonald Observatory, the Lowell Discovery Telescope (LDT), and the Gemini South telescope \citep{mace18}. 

The IGRINS YSO Survey is uniquely suited to determining physical parameters because i) The simultaneous coverage of the H- and K-band (1.45--2.5~$\mu$m) of IGRINS includes separable photospheric and disk contributions to the YSO spectrum. ii) The fixed spectral format of IGRINS provides similar spectral products for each object at each epoch. iii) {Multiple-epoch observations can provide a means to characterize or average over variability.} iv) {The velocity resolution of $\sim$7~km~s$^{-1}$ (R$\sim$45000) }is smaller than the typical young star rotational velocity and can resolve magnetic fields $\gtrsim$~1 kG.

In \cite{lopezv21} (hereafter \citetalias{lopezv21}), we presented the first results of the IGRINS YSO survey. This first study consisted of the determination of photospheric temperature (\teff), surface gravity (\logg), magnetic field (B), projected rotational velocity (\vsini), {and the K-band veiling ($r_K$)} for 110 YSOs located in the Taurus-Auriga star-forming region and 19 young stars in the TW Hydrae association (TWA).
The IGRINS YSO survey analysis continued with the determination of low-resolution veiling spectra for 144 Taurus-Auriga members in \cite{kidder2021} \citepalias{kidder2021}. 
In this paper, the third of the series, we extend our analysis to K- \& M-type YSOs in the Ophiuchus and Upper Scorpius star-forming regions \citep{elias78,lada84,wilking89}. Combining the results of \citetalias{lopezv21}\ with those obtained here improves our understanding of YSO stellar parameters as a function of age.  

%-----------------------
%---------------------------------
\section{Observations and sample}
Among the IGRINS YSO Survey, we included bright objects (K $<$~11 mag) {with spectral types between K0 and M5 \citep{2005AJ....130.1733W, 2006AA...460..695T,2010AA...521A..66R,2020AJ....159..282E,2020AJ....160...44L}} that are { classified as member of Ophiuchus and Upper Scorpius by \cite{2013ApJS..205....5H}, \cite{rebull18}, and  \cite{2020AJ....159..282E}}. 

We observed our sample with IGRINS at the McDonald Observatory 2.7~m telescope, the LDT, and the Gemini South Telescope between 2014 and 2019. We followed the observing and data reduction methods as outlined in \citetalias{lopezv21}. In brief, we observed the YSOs and A0V telluric standards\footnote{An A0V telluric standard star was observed at a similar airmass within two hours before or after the science target.} by nodding the stars along the slit in patterns made up of AB or BA pairs, where A and B are two different positions on the slit. We aimed to obtain a minimum signal-to-noise ratio (SNR) of 70 in the K-band by adjusting the total exposure time based on the IGRINS exposure time calculator\footnote{\url{https://wikis.utexas.edu/display/IGRINS/SNR+Estimates+and+Guidelines}}. 
An SNR above 70 is sufficient to use our methods and to produce reliable stellar parameters. 

We employed the IGRINS pipeline \citep*{lee17}\footnote{\url{https://github.com/igrins/plp/tree/v2.1-alpha.3}}  to reduce all the spectroscopic data. The pipeline produces a telluric corrected spectrum with a wavelength solution derived from OH night sky emission lines and telluric absorption lines. 
The wavelength solution was corrected for the barycenter velocity determined with {\sc zbarycorr} \citep{wright2014}. 

About a third of the YSOs in our sample have more than one epoch available. {For the determination of average parameters (this work) a weighted-average spectrum is produced from multi-epoch observations using the SNR at each data point as the weight. The standard deviation of the mean gives the final uncertainties per data point. Future studies of the IGRINS YSO Survey will look at multi-epoch variability.}

{We identified binaries (or multiples) in the YSO sample with a separation of less than 2 arcseconds. When the separation is greater than 2 arcseconds, the objects can be observed individually by IGRINS. If the separation is less than 2 arcseconds, the resultant IGRINS spectrum will contain some flux from both components.
A complete determination of the binary nature of these stars is beyond the scope of this work. 
Binary searches for the YSOs in our sample have identified some binaries and provided limits on detection methods \citep{1995ApJ...443..625S, 2002AJ....124.2841H, 2003ApJ...584..853P, 2005AA...437..611R, 2007ApJ...657..338P, 2008ApJ...679..762K, 2010ApJ...712..925C, 2010ApJ...709L.114D, 2019ApJ...878...45B}.%, but unresolved binaries can not be definitively ruled out for the objects in our sample. 
We marked known binaries and left as singles those objects that appear as limits in binary classification studies or meet our separation criterion. Those objects that lack information in the literature have a question mark. We have excluded double-lined binaries from our sample through careful inspection of the YSO Survey spectra. The binary status and basic information of our sample are available in Table~\ref{tab:res}. }

%%%%%%%%%%%%%%%%%%%%%%%%%%%%
%%%%%%%%    METODO
%%%%%%%%%%%%%%%%%%%%%%%%%%%%
\section{Stellar parameter determination}
We followed the same methods as in \citetalias{lopezv21}\ to obtain \teff, \logg, B, \vsini, {and $r_K$.} Our method used a Markov chain Monte Carlo (MCMC) analysis as implemented in the code {\it emcee} \citep{emcee}. 
We computed a four-dimensional (\teff, \logg, B, \vsini) grid of synthetic spectra using the {\sc moogstokes} code \citep{deen13}. 

Moogstokes synthesizes the emergent spectrum of a star taking into account the Zeeman broadening produced by the presence of a photospheric magnetic field. In this work, we used the MARCS atmospheric models \citep{marcs} with {solar metallicity (suitable for YSOs; 
\citealp{dorazi11})} and {the astrophysical-inferred modifications to the Vienna Atomic Line Database (VALD3; \citealp{vald3}) line transition data ($\log gf$ and Van der Waals constant values) presented by \cite{flores19}}. We used a micro-turbulence of 1~km~s$^{-1}$ as low-mass stars have micro-turbulence values between 0 and 2~km~s$^{-1}$ \citep[e.g., ][]{gray05,bean06}
 
{Our grid of synthetic spectra matched the IGRINS spectral resolution (R$\sim$45,000) and covers the parameter space as follows: from 3000 to 5000~K in \teff\ 
(steps of 100~K up to 4000~K, and 250~K above 4000 K), from 3.0 to 5.0~dex in 
\logg\ (steps of 0.5~dex), from 0 to 4 kG in B (steps of 0.5~kG), and  values from 2 to 
50~km~s$^{-1}$ in \vsini\ (steps of 2~km~s$^{-1}$). The grid steps\footnote{The grid steps of \teff\ and \logg\ are defined by the model atmosphere, while we selected the steps of B and \vsini.} are spaced enough to differentiate the effects of the parameters on the spectra.}

The synthetic spectral grid covers four spectral intervals in the K-band, namely Na, Ti, Ca, and CO. The intervals include neutral atomic (Fe, Ca, Na, Al, and Ti) and molecular lines (CO) that are sensitive to changes in the different stellar parameters (see Figure~2 of \citetalias{lopezv21}).
The four spectral intervals are shown in Figure~\ref{fig:bestfit}.
Each spectral region was normalized using an interactive Python script. 
The script fits a polynomial of order $n$ to a custom number of flux bins, usually
between 10-20 bins, and a polynomial of order 1-4.
After normalizing the continuum, we carried out the MCMC comparing observed and synthetic spectra in the K-band spectral regions by allowing \teff, \logg, B, \vsini, and veiling to vary along with small continuum ($<$~6\%) and wavelength ($<$~1.0\AA) offsets. In each MCMC trial, we linearly interpolated within the four-dimensional (\teff, \logg, B, \vsini) synthetic spectral grid to obtain the corresponding spectrum with the sampled set of parameters. The interpolated synthetic spectrum was then artificially veiled by the veiling parameter and re-normalized for each region. A single veiling value was used for all the K-band wavelength regions. 

Finally, from the posterior probability distributions of the MCMC, we took the 50th percentile as the most likely value for each stellar parameter. We computed the total uncertainty for each parameter as the quadrature sum of the formal fit errors (the larger of the 16th and 84th percentile of the posterior probability distribution) and the systematic errors. {Systematic errors were determined in \citetalias{lopezv21}\ and we assume the same values for this work. They are 75~K, 0.13~dex, 0.26~kG, and 1.7~km~s$^{-1}$ for \teff, \logg, B, and \vsini, respectively.}

%##########################
%##### Resultados
%##########################
\section{Results}
\begin{figure}
    \centering
    \includegraphics[width=\columnwidth]{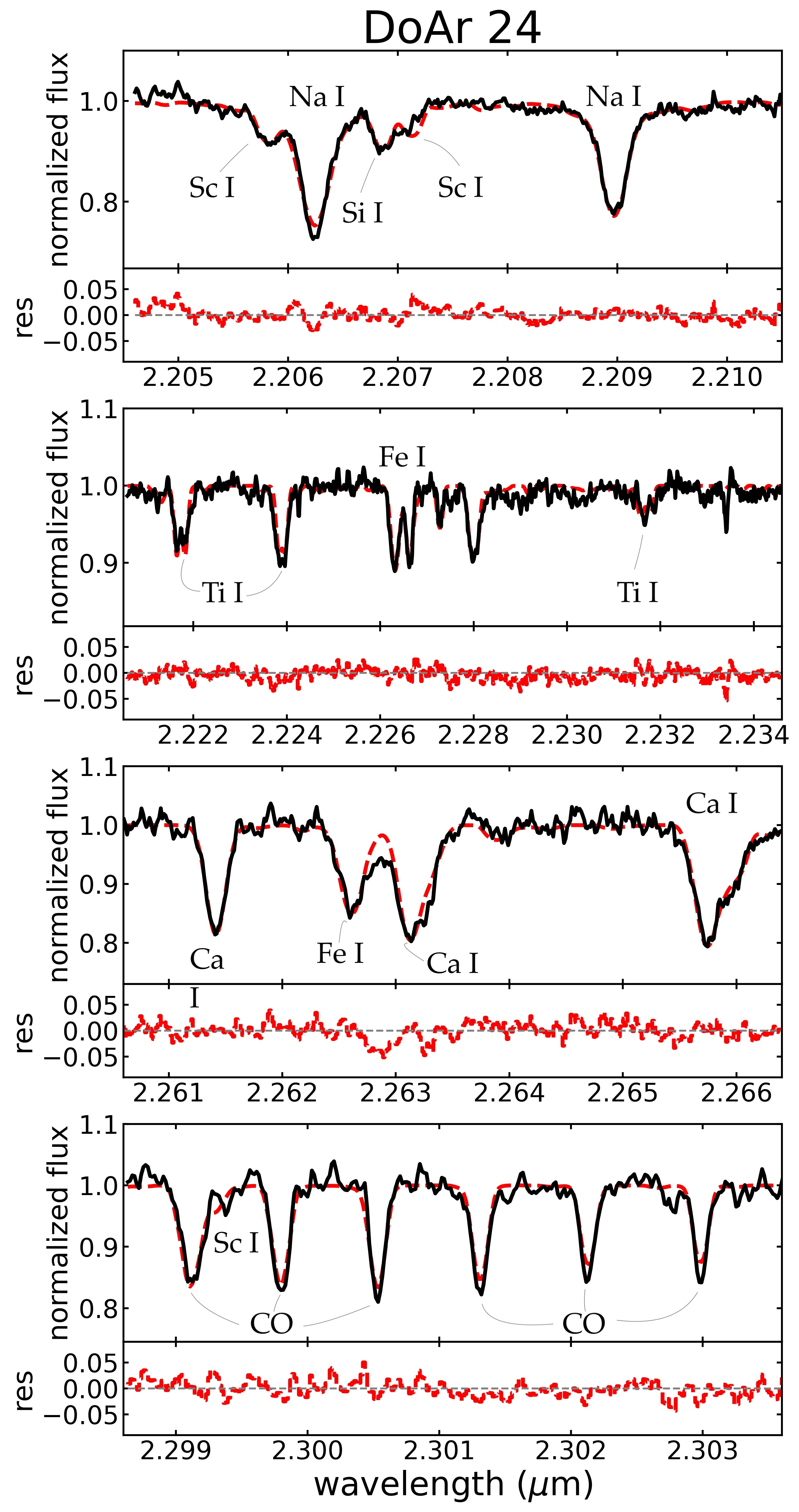}
    \caption{The spectral regions used in this work for the star DoAr 24 (solid black line, 2MASS J16261706-2420216). We included the best fit synthetic spectrum (red dashed line), which has  \teff~=~4092~K, \logg~=~4.11~dex, B~=~2.12~kG, \vsini~=~10.6~km~s$^{-1}$, and a K-band veiling of 0.79. The bottom panels show the residuals between the observed and the synthetic spectra. The level of agreement between the best-fit synthetic spectrum and the DoAr 24 observed spectrum was considered a good determination. We also indicate some atomic and molecular lines present in the spectral regions.}
    \label{fig:bestfit}
\end{figure}

Once we determined the stellar parameters of our sample {(Table~\ref{tab:res})}, we categorized our determinations, based on a visual inspection and $\chi^2$ statistics, with a numerical quality flag equal to 0, 1, or 3 if they are good, acceptable, or poor determinations. 
We identified objects with lower/upper limits outside or at the edge of our grid with a flag of 2. We used each stellar parameter's total uncertainty to compute its lower and upper limits. In Figure~\ref{fig:bestfit}, we show an illustrative example of how the synthetic spectra reproduce the observations. 

We identified {31}, {25}, {5}, and {9} stars whose parameters are good, acceptable, at the edge of the grid or poor determinations, respectively. We excluded poor determinations in our further analysis and referred to the {61} remaining stars as our sample.

\begin{deluxetable*}{llllllllllllllllll}
\tabletypesize{\tiny} 
\rotate
\tablewidth{700pt}
 \tablecaption{Basic information and results for our Ophiuchus and Upper Scorpius sample. The flag column equals 0, 1, 2, and 3 indicates the good, acceptable, at the edge of the grid, and bad determinations of atmospheric parameters. The K magnitude is from \citet{cutri03}.  The full version of this table is available in the online version of the paper.\label{tab:res}}
\tablehead{
\colhead{2MASS} & \colhead{Name} & \colhead{N} & \colhead{K} & \colhead{SpT} & \colhead{ref} & \colhead{cluster} & \colhead{ref} & \colhead{Bin} & \colhead{ref} & \colhead{\ext} & \colhead{$\alpha_{K_s-W4}$} & \colhead{flag} & \colhead{\teff} & \colhead{\logg} & \colhead{B} & \colhead{\vsini} & \colhead{$r_{\rm K}$}\\
\colhead{} & & \colhead{} & \colhead{mag} & \colhead{} & \colhead{} & \colhead{} & \colhead{} & \colhead{} & \colhead{} & \colhead{mag} & \colhead{} & \colhead{} & \colhead{K} & \colhead{dex} & \colhead{kG} & \colhead{km$s^{-1}$} & \colhead{}} 
\startdata
J16081474-1908327 & EPIC 205152548 & 2 & 8.4 & K2 & 18 & USco & 14 & Y & 16 & 0.6$\pm$0.1 & -2.79$\pm$0.06 & 1 & 4652$\pm$172 & 4.55$\pm$0.34 & 1.05$\pm$0.70 & 8.4$\pm$3.8 & 0.31$\pm$0.15\\
J16082324-1930009 & EPIC 205080616 & 1 & 9.5 & K9 & 18 & USco & 14 & N & 8 & 0.8$\pm$0.5 & -1.09$\pm$0.22 & 1 & 3782$\pm$277 & 4.16$\pm$0.44 & 2.03$\pm$0.91 & 14.3$\pm$4.1 & 0.15$\pm$0.11\\
J16090075-1908526 & EPIC 205151387 & 2 & 9.2 & M1 & 18 & USco & 14 & N & 8 & 0.7$\pm$0.5 & -0.70$\pm$0.27 & 0 & 3636$\pm$188 & 4.22$\pm$0.32 & 1.88$\pm$0.45 & 8.8$\pm$2.9 & 0.26$\pm$0.09\\
J16093030-2104589 & RX J1609.5-2105 & 1 & 8.9 & M0 & 18 & USco & 15 & N & 16 & 0.6$\pm$0.6 & -2.75$\pm$0.04 & 0 & 4021$\pm$278 & 3.98$\pm$0.47 & 1.44$\pm$0.90 & 10.7$\pm$3.9 & 0.20$\pm$0.11\\
J16110890-1904468 & ScoPMS 44 & 4 & 7.7 & K4 & 18 & USco & 15 & N & 16 & 1.4$\pm$0.2 & -2.78$\pm$0.07 & 1 & 4220$\pm$160 & 4.50$\pm$0.33 & $<$2.66 & 27.1$\pm$3.4 & 0.44$\pm$0.09\\
J16113134-1838259 & V* V866 Sco & 10 & 5.8 & K5 & 18 & USco & 14 & Y & 3 & 3.4$\pm$0.5 & -0.49$\pm$0.05 & 1 & 3919$\pm$230 & 3.49$\pm$0.39 & $<$1.52 & 16.5$\pm$3.6 & 7.45$\pm$1.00\\
J16142029-1906481 & EPIC 205158239 & 1 & 7.8 & M0 & 18 & USco & 14 & N & 13 & 2.8$\pm$0.8 & -0.71$\pm$0.07 & 1 & 3911$\pm$391 & 4.13$\pm$0.60 & $<$1.97 & 24.5$\pm$8.3 & 1.64$\pm$0.53\\
J16153456-2242421 & V* VV Sco & 1 & 7.9 & M0 & 18 & USco & 15 & Y & 8 & 1.2$\pm$0.7 & -0.92$\pm$0.22 & 1 & 3685$\pm$356 & 4.17$\pm$0.56 & 2.47$\pm$1.05 & 10.2$\pm$5.9 & 1.16$\pm$0.31\\
J16211848-2254578 & EPIC 204290918 & 1 & 10.2 & M2 & 18 & USco & 14 & N & 11 & 2.1$\pm$0.7 & -0.84$\pm$0.16 & 0 & 3414$\pm$229 & 4.09$\pm$0.45 & $<$1.51 & 19.4$\pm$3.4 & 0.67$\pm$0.16\\
J16220961-1953005 & EPIC 205000676 & 1 & 8.9 & M3.7 & 17 & USco & 14 & Y & 16 & 1.6$\pm$0.7 & -2.08$\pm$0.09 & 1 & 3371$\pm$228 & 3.80$\pm$0.40 & $<$1.24 & 16.1$\pm$2.9 & 0.19$\pm$0.11\\
J16232454-1717270 & EPIC 205483258 & 3 & 9.7 & M2.5 & 18 & USco & 15 & ? & -- & -- & -- & 3 & -- & -- & -- & -- & -- \\
\enddata
\tablereferences{
(1) \citet{1995ApJ...443..625S}
(2) \citet{2002AJ....124.2841H}
(3) \citet{2003ApJ...584..853P}
(4) \citet{2005AA...437..611R}
(5) \citet{2005AJ....130.1733W}
(6) \citet{2006AA...460..695T}
(7) \citet{2007ApJ...657..338P}
(8) \citet{2008ApJ...679..762K}
(9) \citet{2010AA...521A..66R}
(10) \citet{2010ApJ...709L.114D}
(11) \citet{2010ApJ...712..925C}
(12) \citet{2013ApJS..205....5H}
(13) \citet{2014ApJ...785...47L}
(14) \citet{rebull18}
(15) \citet{2018AJ....156...76L}
(16) \citet{2019ApJ...878...45B}
(17) \citet{2020AJ....159..282E}
(18) \citet{2020AJ....160...44L}
}
\end{deluxetable*}

%------------------
%------------------------
%-------------------------------------
\subsection{Photospheric temperature}
The derived temperatures for non-binary Ophiuchus and Upper Scorpius YSOs follow the same trend with spectral type as was found for single Taurus YSOs in \citetalias{lopezv21}. 
To improve our previous IGRINS temperature scale, we combined our previous \teff\ values of Taurus and TWA (\citetalias{lopezv21}) with the \teff\ determinations made in this work. The combined sample (\citetalias{lopezv21}\ and this work) comprises {190} YSOs.

We computed the mean \teff\ and the standard deviation of the mean in bins of $\pm$0.5 spectral type subclasses to construct the IGRINS temperature scale. We included in our IGRINS temperature scale spectral type bins with more than one object.

\begin{figure}
    \centering
    \includegraphics[width=\columnwidth]{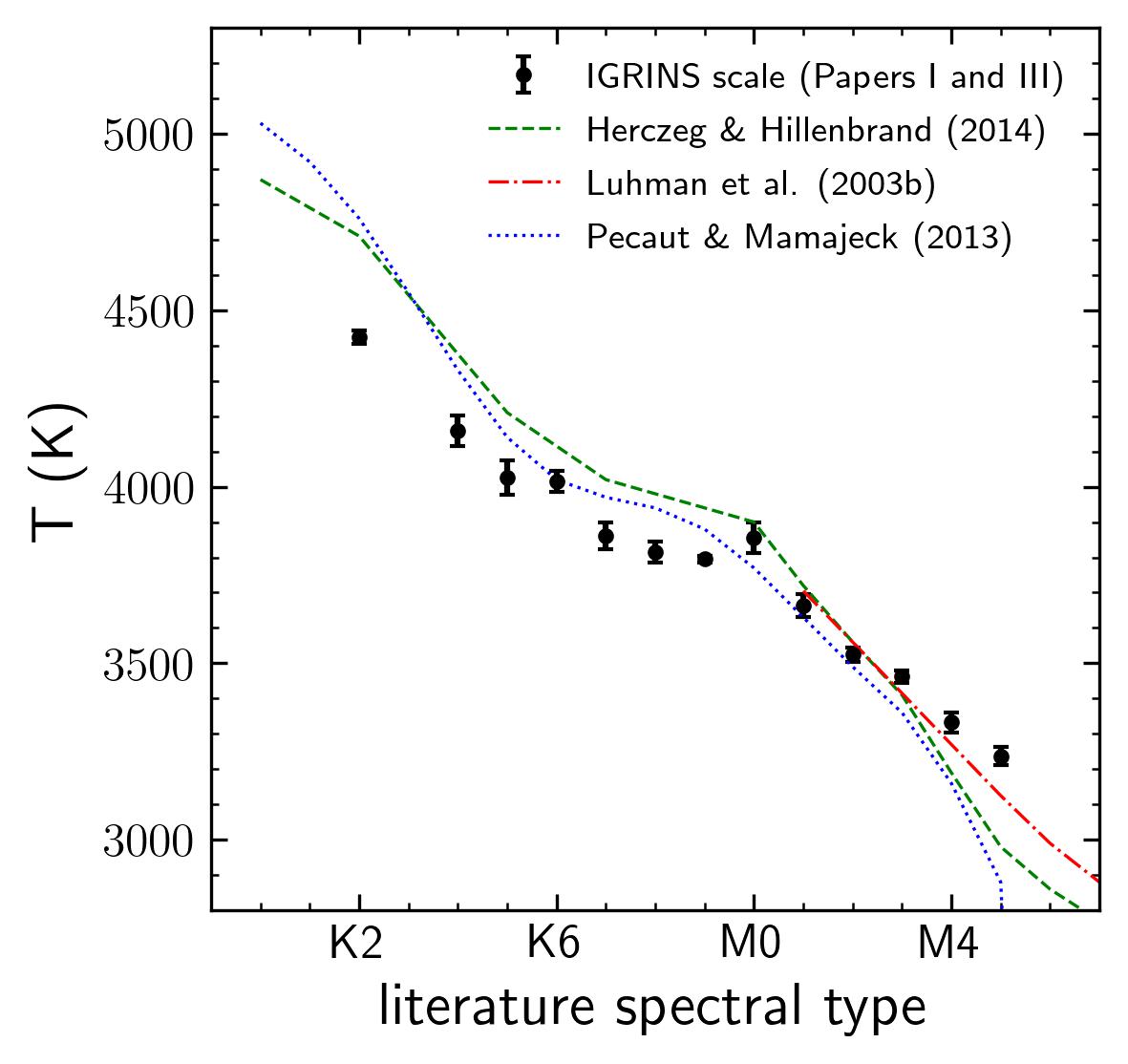}
    \caption{Temperature as a function of the literature spectral type. The dots represent the IGRINS temperature scale, determined using the temperatures obtained for the non-binary YSOs of Ophiuchus, Upper Scorpius, TWA, and Taurus. The different lines represent the temperature scales of \citet{herczeg14}, \citet{luhman03b} and \citet{pecaut13}. Our temperature scale agrees with the published temperature scales for M0-M3 stars but follows the \cite{luhman03b} scale for late types. However, our temperatures are cooler than the published scales in K stars. The error bar in our temperature is the standard deviation of the mean within the $\pm$0.5 spectral type sub-classes.}
    \label{fig:temp}
\end{figure}

In Figure~\ref{fig:temp} we compare the updated IGRINS temperature scale, which is in Table~\ref{tab:scale}, to the published scales of \cite{pecaut13}, \cite{herczeg14} and \cite{luhman03b}. In general, the shape of our temperature scale is similar to the published ones. The \teff\ values for M0-M3 stars agree with the three temperature scales. However, our scale agrees better with the \cite{luhman03b} for late spectral types. For the K stars, we found that our temperature values are cooler than the temperature scales of \cite{pecaut13} and \cite{herczeg14}. 

The differences between our \teff\ values and the literature could be an effect of having a small sample of K stars. Also, it can be explained as a cumulative result of using different atmospheric models and line lists, different methodologies (spectroscopy vs. photometry), different wavelength intervals (optical vs. infrared), the effect of B field on \teff, and even differing SpT classifications. 

For example, typical SpT uncertainties are about one subclass ($\sim$ 100 K); however, in many cases, the SpT determined by different groups for a certain YSO disagrees by more than two subclasses, even more than five in extreme cases. Additionally, omitting the B field in the \teff\ determination produces temperatures, on average, $\sim$40--70K cooler than if the B is considered (\citetalias{lopezv21}). 

One way that we reduce SpT uncertainties in this work is to use homogeneous SpT classifications.
The SpT of the Taurus sample comes from \cite{luhman17}, while for Ophiuchus and Upper Scorpius members, the SpT comes mostly from \cite{2020AJ....159..282E} and \cite{2020AJ....160...44L}. 
These three studies measured SpT by comparing absorption bands (TiO, VO, Na I, K I, and H2O) between observed and standard optical and infrared spectra. Their reported uncertainties in SpT are $\pm$0.25 and $\pm$0.5 subclasses for optical and infrared types.

\begin{deluxetable}{lccclc}
\tabletypesize{\normalsize}
\tablewidth{0pt}
 \tablecaption{Temperature scale determined using IGRINS data along with those of \citet[][L03]{luhman03b}, \citet[][PM13]{pecaut13}, and \citet[][HH14]{herczeg14}.This IGRINS temperature scale is an update of that presented in Paper I. The last column correspond to the number of objects with which we computed the mean \teff\ value and the standard error of the mean within a $\pm$0.5 spectral type subclasses. We only included spectral type bins with more than one determination. \label{tab:scale}}
\tablehead{ \colhead{SpT} & \colhead{L03} & \colhead{PM13} & \colhead{HH14} & \colhead{IGRINS} & \colhead{N} \\
& \colhead{(K)}& \colhead{(K)} & \colhead{(K)} & \colhead{(K)} & }
\startdata 
K0  & --    &  5030  & 4870  &  --  & -- \\
K1  & --    &  4920  & --  &  -- & -- \\
K2  & --    &  4760  & 4710  &  4424$\pm$18  & 2 \\
K3  & --    &  4550  & --  &   --& -- \\
K4  & --    &  4330  & --  &   4158$\pm$43 & 2 \\
K5  & --    &  4140  & 4210  &  4025$\pm$49  & 5 \\
K6  & --    &  4020  & --  &   4016$\pm$30& 14 \\
K7  & --    &  3970  & 4020  &  3861$\pm$37   & 4 \\
K8  & --    &  3940  & --  &   3815$\pm$29& 5 \\
K9  & --    &  3880  & --  &   3795$\pm$9& 2 \\
M0  & --    &  3770  & 3900  &  3856$\pm$43  & 15 \\
M1  & 3705  &  3630  & 3720  &  3663$\pm$33  & 16 \\
M2  & 3560  &  3490  & 3560  &  3525$\pm$20  & 16 \\
M3  & 3415  &  3360  & 3410  &  3462$\pm$18   & 19 \\
M4  & 3270  &  3160  & 3190  &  3332$\pm$29   & 16 \\
M5  & 3125  &  2880  & 2980  &  3237$\pm$26   & 11
\enddata
\end{deluxetable}

%===========================
\subsection{Surface gravity}\label{sect:gra}
\begin{figure}
    \centering
    \includegraphics[width=\columnwidth]{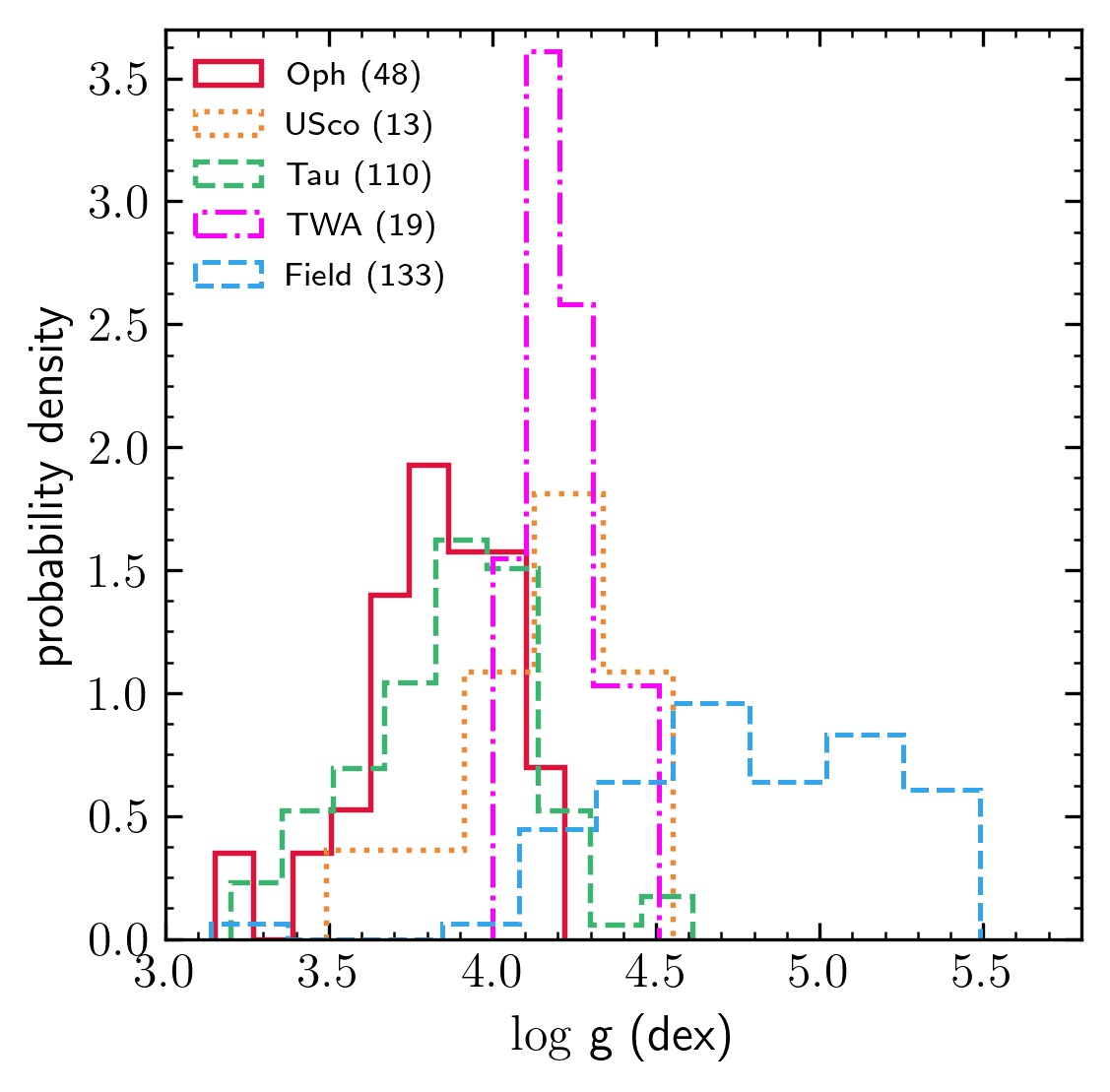}
    \caption{Probability density of \logg\ for 48, 13, 110, 19, YSOs members of Ophiuchus, Taurus, Upper Scorpius, and TWA. We also show the \logg\ determinations obtained for 133 K and M field stars.}
    \label{fig:gdist}
\end{figure}

The \logg\ distributions of Ophiuchus and Upper Scorpius are differentiated in Figure \ref{fig:gdist}. Their Kolmogorov-Smirnov probability\footnote{The Kolmogorov-Smirnov (KS) test computes the probability that two populations come from the same parent distribution.} of {0.1\%} also shows that they are statistically different. The Ophiuchus distribution has a mean \logg\ (and error on the mean) of {3.83$\pm$0.03~dex}, while Upper Scorpius has {4.13$\pm$0.08~dex}. 

Considering the \logg\ as a proxy for age, we can speculate that such differences in \logg\ between samples might be due to age. To test this, we used the combined sample, and we also included the \logg\ values determined for {133} field stars (see Appendix \ref{app:field}) as a more evolved counterpart. 

From Figure \ref{fig:gdist}, we can see that the \logg\ distributions of Taurus and Ophiuchus are very similar as its KS probability of {78\%} also confirms. We also found that the \logg\ distribution of Upper Scorpius is comparable to the TWA distribution (KS probability of {35\%}). 

We also see in Figure \ref{fig:gdist} that the \logg\ distribution of field stars is distinguishable from the YSOs, as is expected for a more evolved sample. There is some overlapping between the low \logg\ values of the field stars, which is about 4.0 dex, and the higher \logg\ determinations of Taurus, Ophiuchus, Upper Scorpius, and TWA.  

{Based on our \logg\ determinations and using them as an age proxy, we can conclude that Ophiuchus and Taurus have a similar age, which is younger than the indistinguishable Upper Scorpius and TWA samples.} %Finally, the K stars are older than our YSOs samples, but younger than the M field stars.}

\begin{figure*}[ht]
    \centering
    \includegraphics[width=\textwidth]{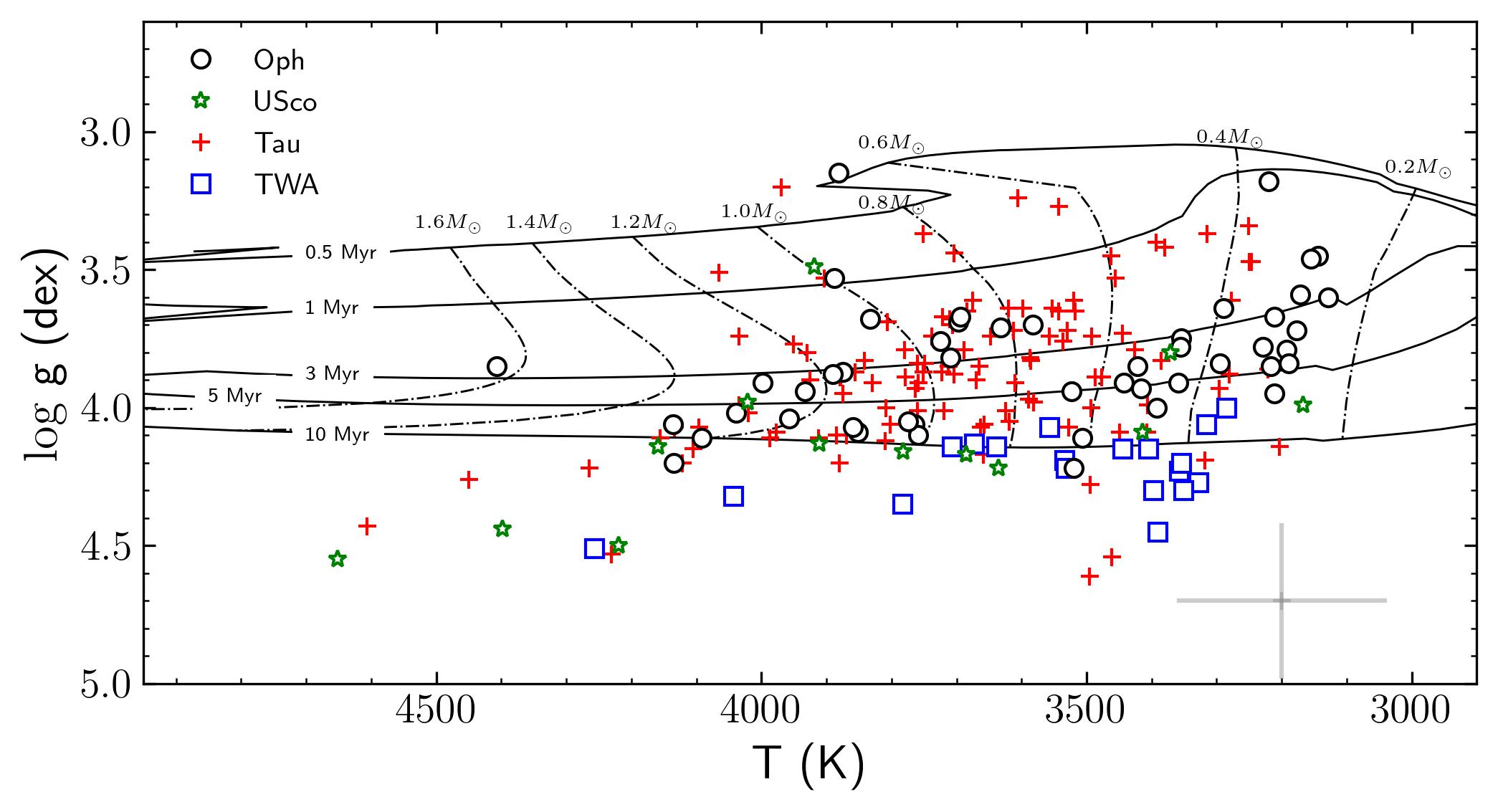}
    \caption{Spectroscopic Hertzprung-Russell or Kiel diagram for 48 Ophiuchus (circles), 110 Taurus-Auriga (plus), 13 Upper Scorpius (stars), and 19 TWA (squares) YSOs. The Taurus and Ophiuchus samples seem to populate similar parts of this diagram, and it is no clear difference in \logg. Some objects of Upper Scorpius and Taurus have a \logg\ value relatively high compared with most of our determinations. 
    {We included magnetic evolutionary tracks of \citet{feiden16}}. Finally, the error bar in the right-bottom corner represents the median uncertainties of our combined sample. \label{fig:shrd}}
\end{figure*}

{In Figure~\ref{fig:shrd}, we produce the Kiel (\teff\ vs. \logg) diagram for the YSOs in Taurus, Ophiuchus, Upper Scorpius, and TWA. We also included in Figure~\ref{fig:shrd} the magnetic evolutionary models of \cite{feiden16} as a theoretical comparison. The Taurus and Ophiuchus objects seem to populate similar parts of the Kiel diagram, and there is no clear difference in \logg\ between them. These objects also have a larger \logg\ dispersion than their counterparts in Upper Scorpius or TWA.}

%=====================
%========================
%================
\subsection{Spectral indices and classification}\label{sect:alpha}
Based on the shape of their de-reddened spectral energy distributions, \cite{lada84} divided the YSOs into three different morphological classes ({\sc i}, {\sc ii}, and {\sc iii}), by means of the spectral index $\alpha = d\log (\lambda F_\lambda) / {d\log \lambda}$.

The class scheme represents an evolutionary sequence of low-mass stars, and it is helpful to understand the star formation process.
Although most of our YSOs have a previous classification, we computed de-reddened $\alpha$ indices for our YSOs to investigate trends with stellar parameters. 

We first constructed the infrared spectral energy distribution (SED) from the {\tt 2MASS} ($J$, $H$, and $K_s$) and {\tt WISE} ($W1$, $W2$, $W3$, and $W4$) magnitudes reported in the {\tt ALLWISE} catalog \citep{cutri13}. We included those magnitudes with a flux signal-to-noise ratio greater than two and ignored the upper limits. 

{We then corrected the observed SEDs by reddening. We used a visual extinction (\ext) estimation that we converted into a extinction in the $K_{\rm s}$ band ($A_{K_{\rm s}}$) by means of $A_{K_{\rm s}}$/A$_{\rm V}$ = 0.112 \citep{rieke85,groschedl19}. We then employed the zero magnitude flux densities, and extinction laws reported in \cite{indebe05} and  \cite{groschedl19} to compute the de-reddened $J, H, K, W1, W2, W3$, and $W4$ fluxes. 
Using the de-reddened fluxes and the central wavelengths of each filter, we created the extinction-corrected infrared SED for those YSOs in our combined sample with \ext estimations. 
We obtained \ext\ values with the code \texttt{MassAge} (Hern\'andez et al. in prep). \texttt{MassAge} {input includes \teff}, Gaia EDR3 (Gp, Rp, Bp), and 2MASS (J and H) photometry to compute \ext\ by minimizing the differences between the observed and expected intrinsic colors of \cite{2020AJ....160...44L} affected by reddening. The reddening \ext\ is changed until the best comparison is found using the minimal $\chi^2$ method.

The uncertainties in \ext\ values are obtained using the Monte Carlo method of error propagation \citep{Anderson76}, assuming Gaussian distributions for the uncertainties in the input parameters. We found photometric data for only {171} objects, for which we computed the \ext\ (see Table \ref{tab:res}). 

{Continuum veiling ($r_{\rm K}$) and \ext\ affect the stellar light similarly, i.e., they redden it. Since we determine \teff\ by taking into account $r_k$, we ensure that MassAge is correctly determining the \ext\ values. Moreover, if we compare the $r_k$ values as a function of their MassAge \ext\, we do not find any trend. Furthermore, these data have a correlation coefficient of 0.18, suggesting a weak correlation between our $r_k$ and \ext\ values. }

The presence, structure, accretion status, and evolutionary state of a circumstellar disk impact the photometric colors, producing an incorrect \ext\ value. As some of our YSOs might still possess a disk, we ran \texttt{MassAge} a second time. This time we did not include the Gaia Bp magnitude, which should be the photometric band most affected by the accretion, to assess the impact of accretion in our sample. 
We found that the \ext\ obtained using all the photometric bands are higher, on average 0.14 mag, than those values determined ignoring Bp magnitude. This difference is lower than the mean uncertainty determined for \ext, which is about 0.4 mag. 

}

Finally, to obtain the infrared spectral index ($\alpha_{K_s-W4}$) we perform a linear fit between the $K_{\rm s}$ and $W4$ de-reddened fluxes. Our infrared indices are reported in Tables~\ref{tab:res} and \ref{tab:tau}.

\begin{figure} 
    \centering
    \includegraphics[width=\columnwidth]{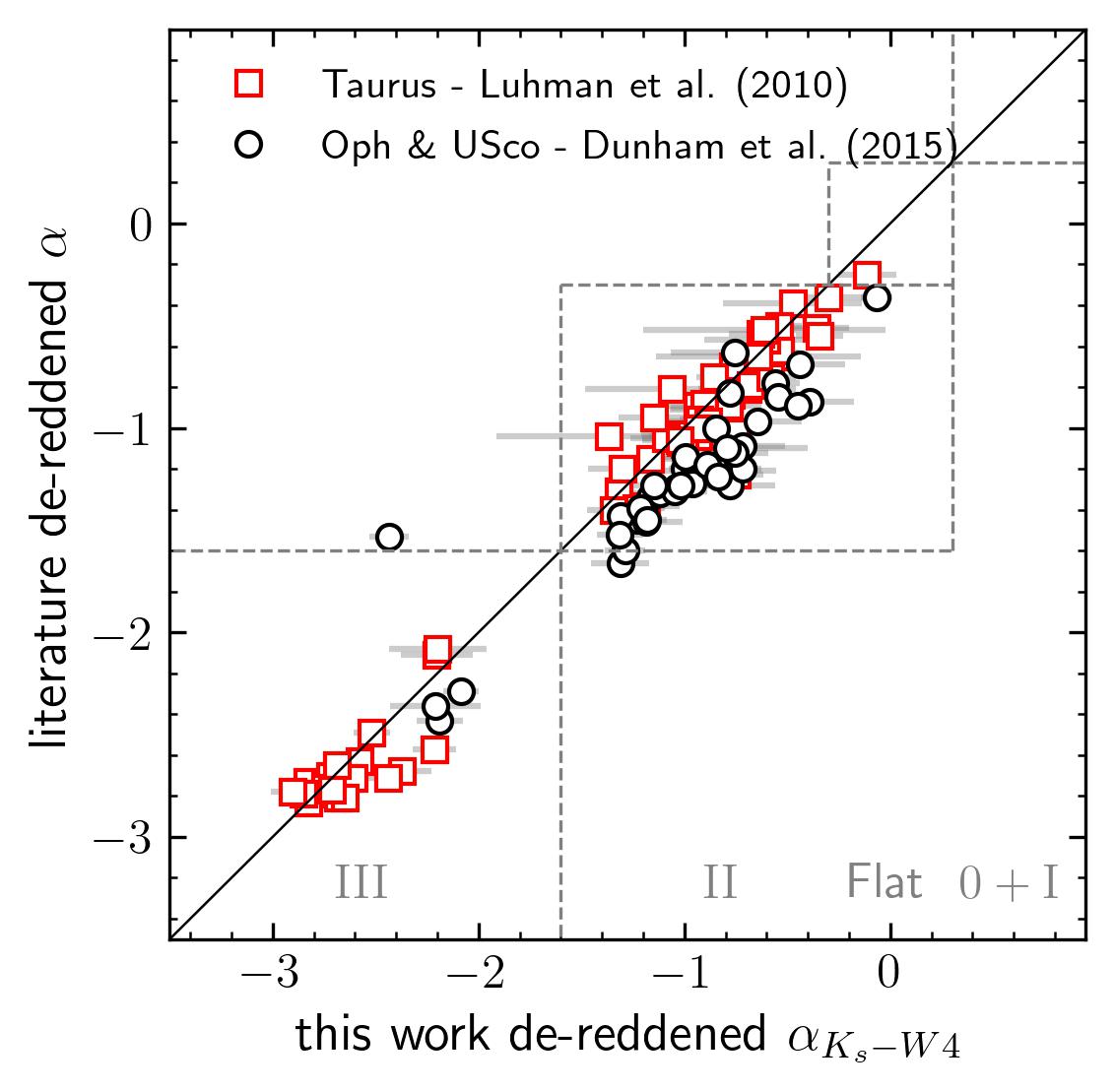}
    \caption{Comparison of the de-reddened $\alpha$ indices determined in this work with those determined in \citet{dunham15} and \citet{luhman10}. The gray squares represent the area where they agree on the class status. We used the classification thresholds of \cite{groschedl19} for Class~{\sc iii}, Class~{\sc ii}, Flat spectrum, and Class~0 + {\sc i}, respectively. The error bar on our determinations is the uncertainty in the best-fit slope. Our indices are in good agreement for Taurus, but they present a systematic offset toward higher values (on average, +0.25) for Ophichus YSOs. The reason for this offset is unclear but might be related to using different photometric data, zero points, and extinction laws.}
    \label{fig:comp_alp}
\end{figure}

As a quality check, in Figure~\ref{fig:comp_alp},  we compared our spectral indices with those determined by \cite{dunham15} and  \cite{luhman10} for YSOs in the Ophiuchus and Taurus star-forming regions, respectively. \cite{dunham15} determined the spectral index $\alpha$ through a linear least-squares fit of all available 2MASS and Spitzer photometry between 2 and 24~$\mu$m. On the other hand, \cite{luhman10} computed the spectral index between four pairs of photometric bands, including the pair $K_{s}$ and Spitzer 24~$\mu$m. These spectral indices are compatible with those determined here, as we obtained them from the same wavelength range (2 - 24~$\mu$m). The only difference between the works of \cite{dunham15}, \cite{luhman10}, and ours is that we used WISE photometry instead of Spitzer.  

Our classifications are consistent with the literature. For Taurus, we found that $\alpha_{K_s-W4}$ values are around the one-to-one line for the entire interval. On the other hand, our Ophiuchus $\alpha$ indices are higher than those determined by \cite{dunham15}, on average, $\sim$0.25. The source of this discrepancy is unclear but is likely related to the difference between the WISE and Spitzer photometry {and the use of different photometric zero points and extinction laws}. 

%----------------
%------------------------
%-----------
\subsection{Spectral indices vs stellar parameters}
Once we homogeneously computed \alfa\ values, we looked for trends between them and the determined stellar parameters. As we mentioned before, the $\alpha$ index informs us of the evolutionary status of the YSO.  

We did not find any correlation between $\alpha_{K_s-W4}$ and \teff. The dispersion of the \teff\ values in the four populations is very similar,  and also between Class ~{\sc ii} and Class ~{\sc iii} YSOs\footnote{We used the classification thresholds of \cite{groschedl19} and our \alfa\ values. Class {\sc ii}: -1.6 $< \alpha <$-0.3; Class {\sc iii}: $\alpha\ \leq$ -1.6 .}. Regarding \logg, we found that, on average, \logg\ values of Class ~{\sc iii} are higher than their Class ~{\sc ii} counterparts. We found a mean and standard deviation of the mean for Class~{\sc ii} YSOs of {3.86$\pm$0.02~dex}, while for Class~{\sc iii} we found {4.02$\pm$0.04~dex}. 

Similarly, we found that \vsini\ values of Class~{\sc iii} are slightly higher than that of Class~{\sc ii}. We found a mean value of {15.0$\pm$0.7~km~s$^{-1}$} and {18.2$\pm$1.4~km~s$^{-1}$} for the Class~{\sc ii} and Class~{\sc iii} YSOs, respectively. This result can be related to the disk locking scenario in which the presence of an accretion disk regulates the stellar rotation \citep{serna21}.

We consider 65 B determinations from Taurus, Ophiuchus, Upper Scorpius, and TWA. We did not find any trend with the $\alpha_{K_s-W4}$ values and the B-field. Both Class ~{\sc ii} and Class ~{\sc iii} YSOs have B fields very similar between them. We found a mean value of {1.96$\pm$0.06~kG} for Class~{\sc ii} and {2.04$\pm$0.15~kG} for Class~{\sc iii} YSOs, respectively. 

\begin{figure}
    \centering
    \includegraphics[width=\columnwidth]{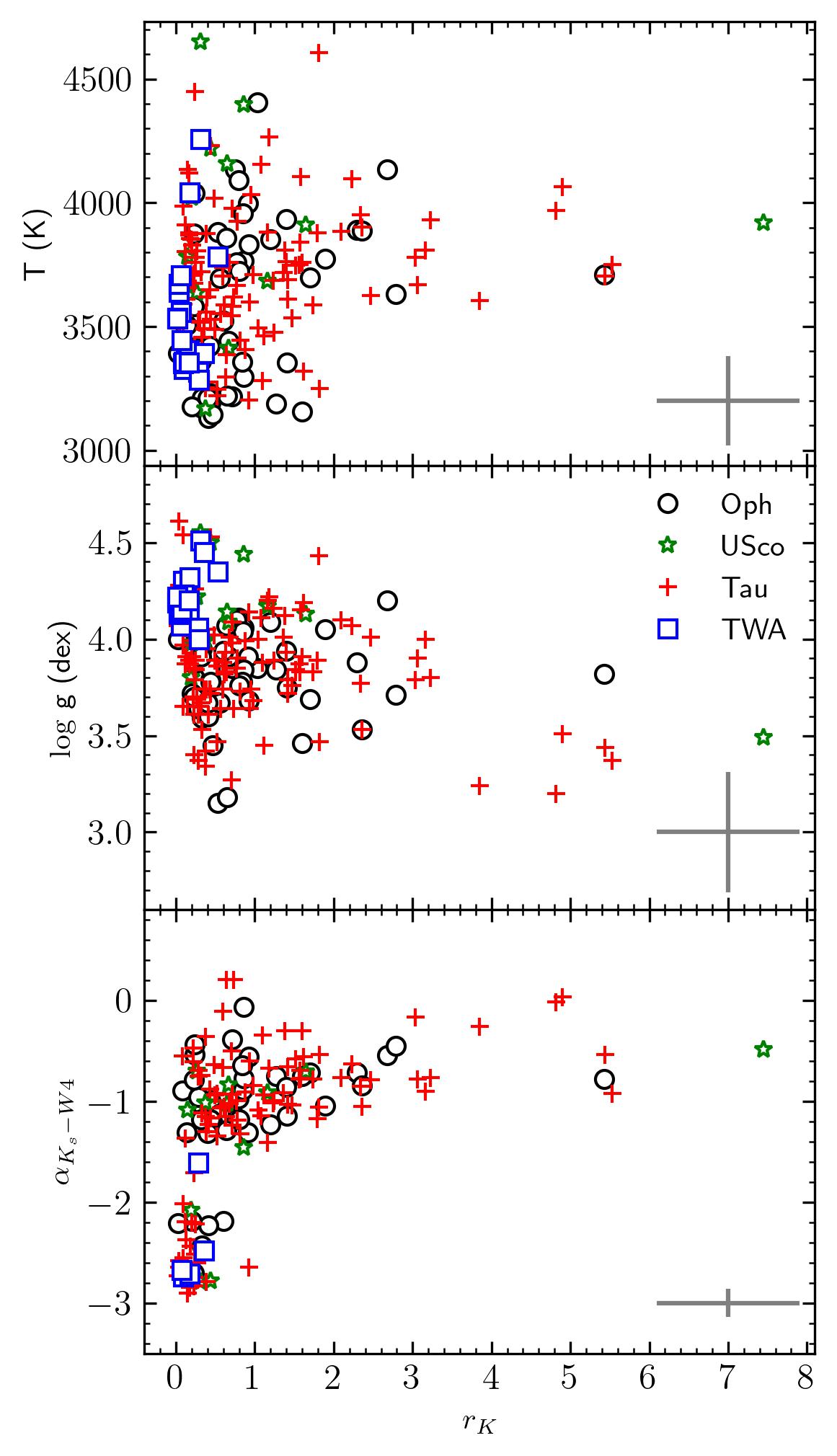}
    \caption{Comparison between the K-band veiling and \teff\ (top), \logg\ (middle), and $\alpha_{K_s-W4}$ (bottom) for the Ophiuchus (black circles), Upper Scorpius (green stars), Taurus (red crosses), and TWA (blue squares). The veiling values of Taurus, Ophiuchus, and Upper Scorpius are more spread out than those of TWA. We also included in each panel the mean error bar}.
    \label{fig:veil}  
\end{figure}

The strongest and expected trend of $\alpha_{K_s-W4}$ is seen for the K-band veiling values {(see Figure \ref{fig:veil})}. More evolved objects have lower K-band veiling values ($r_{\rm K}$ $\lesssim$ 0.6). Interestingly, the K-band veiling values show a large dispersion for Class ~{\sc ii} YSOs. Such a distribution could be related to the variable nature of the Class ~{\sc ii} YSOs or intrinsic differences between the YSOs.

We also compared \teff\ and \logg\ with the K-band veiling in Figure~\ref{fig:veil}. We can see that the TWA objects have a K-band veiling less than $\sim$ 0.8. The veiling values of Taurus, Ophiuchus, and Upper Scorpius are more spread out than those of TWA. Most are between 0 and 3.2, while a few YSOs have K-band veiling greater than 3.2 and up to $\sim$ 7.3. 

{Although of low significance, there seems to be a trend toward higher \rk\ as the \logg\ decreases. The latter suggestion implies that less evolved objects have high veiling values., which aligns with the idea that the veiling is related to the presence of a circumstellar disk. Also, this behavior might be due in part to a degeneracy between $r_{\rm K}$ and \logg, as their effects on the spectral lines are similar.}

In terms of \teff, between $\sim$3600~K and $\sim$4100~K, the \rk\ could be as high as $\sim$7.3. This differs from other temperature ranges, where the highest K-band veiling is around 2.

%---------------------
%--------------------------
%==========================
\subsection{Magnetic field}\label{sect:magne}
The ability to determine the B field from the Zeeman broadening is related directly to the spectral resolution and the \vsini\ of the star. Based on the results of \cite{hussaini20}, the minimum B field we can detect {from IGRINS spectra} should be greater than 1.0 kG, and this scales with the \vsini\ as {B $\geq$ \vsini/8 (kG/km s$^{-1}$)}. We consider B values that do not meet this criterion to be a non-detection. We found that 7 and 20 B field values meet the detection threshold in the Upper Scorpius and Ophiuchus star-forming regions, respectively. The Kolmogorov-Smirnov probability computed between the two samples of B field determinations is 50\%, which means that both samples are not statistically different at that level. In Table~\ref{tab:res} we report both determinations and limits for the B field of our targets.

{The combined sample (\citetalias{lopezv21} and this work) has 78 B-field determinations (20 in Ophiuchus, 7 in Upper Scorpius, 41 in Taurus, and 10 in TWA) and 112 B-field limits. We have compared stellar parameters for the YSOs with B-field detections and non-detections looking for possible trends.  
In terms of \teff, we found that the mean \teff\ for the B-field determinations and B-field limits is very similar, {3675}~K and 3655 K, respectively. This result suggests that the two groups are indistinguishable in terms of \teff, as it is also confirmed by their KS test probability of {35\%}. 
In terms of K-band veiling, we found a KS probability of {5\%} with a mean K-band veiling for the B-field determinations of {0.88} and the B-field non-detections of {0.92}.}
{We then focused on the B-field determinations and split our 78 determinations into two B-field strength bins: low B-field ($1 \le$ B $ < 2.0$ kG) and high B-field (B$\ge$2.0 kG)}. 
%We then focus on the B-field strength and its relation to the stellar parameters. We divided the B-field determinations sample into two bins: . 
According to the KS probability, the low and high B-field distributions are statistically the same for \teff\ and K-band veiling at {40\%} and {14\%}, respectively.

Interestingly, we found a mean \logg\ of {3.88} dex and {4.08} dex for the low B- and high B-field bins, respectively. We also found a KS probability of about {1\%}, meaning that both strength bins are different in terms of \logg. {This is the expected behavior if most of the B-field comes from primordial field conservation during radius contraction of pre-main-sequence stars \citep{moss03}.}

%---------------------------------
%----------------------------------
%---------------------------------
\subsection{Temperature bins}{
The homogeneous stellar parameter determination that we perform for the YSOs in Ophiuchus,  Upper Scorpius, Taurus, and TWA, along with the field stars, allowed us to analyze the similarities and differences of these samples from an evolutionary point of view.

As we see in the previous sections, there are differences between the YSO populations in their \logg\ distributions, which we attribute to differences in age. However, the stellar mass plays a crucial parameter in the evolution of the stars. \teff\ is a rough proxy for stellar mass in most evolutionary models, as constant mass tracks are primarily vertical in the Kiel diagram (e.g., Figure \ref{fig:shrd}).

For that reason, and to fully understand if the differences found in \logg\ are due to age, we separated our star-forming regions into three different \teff\ bins. Our YSO \teff\ values span a range of $\sim$1500 K, between $\sim$3100 and $\sim$4600 K, so our bins comprise stars within 500 K. We name our bins as low (3100 $\le$ \teff\ $<$ 3600 K), med (3600 $\le$ \teff\ $<$ 4100 K), and high (\teff\ $\geq$ 4100 K) temperature, and they contains 82, 92, and 16 objects, respectively. 
In Figure \ref{fig:mass_bin}, we show how the \logg, \vsini, and K-band veiling behave in each mass bin and each star-forming region. We also include field stars in the \logg\ and \vsini\ analysis.

In the four star-forming regions, the behavior with \logg\ (left panel Figure \ref{fig:mass_bin}) is the same. Objects in the low and med \teff\ bins have lower mean \logg\ values than those in the high \teff\ bin. However, in the field stars, the high \teff\ bin has low mean \logg\ values. According to the evolutionary tracks of \cite{marigo17}, for an age of $\sim$ 500 Myr, stars with \teff\ between 3000 and 4000 K (roughly our low and med \teff\ bins) have \logg\ values ranging from $\sim$ 5.0 to 4.7 dex, while stars with K spectral types (high \teff\ bin) have \logg\ in the range of $\sim$4.66 -- 4.63 dex. Our \logg\ values agree with the theoretical predictions of \cite{marigo17} for the M stars. However, our results for the K stars are lower by $\sim$0.3 dex than such predictions. Considering that \logg\ is a challenging parameter to determine, we need a more detailed analysis of the field star sample to address the source of this \logg\ discrepancy in the K stars, which is beyond the scope of this paper.
Our \logg\ values follow the trend of the theoretical predictions where K stars have higher \logg\ than M stars, and they are helpful in the relative sense we have used them throughout the paper. 
We see from the middle panel of Figure \ref{fig:mass_bin} that the mean \vsini\ values create a descendent ladder of the mean \vsini\ values in the following order: Taurus, Ophiuchus, Upper Scorpius, TWA, and field stars.
Finally, the mean values of the K-band veiling (right panel Figure \ref{fig:mass_bin}) are very similar between \teff\ bins and between star-forming regions.
\begin{figure*}[!h]
    \centering
    \includegraphics[width=\textwidth]{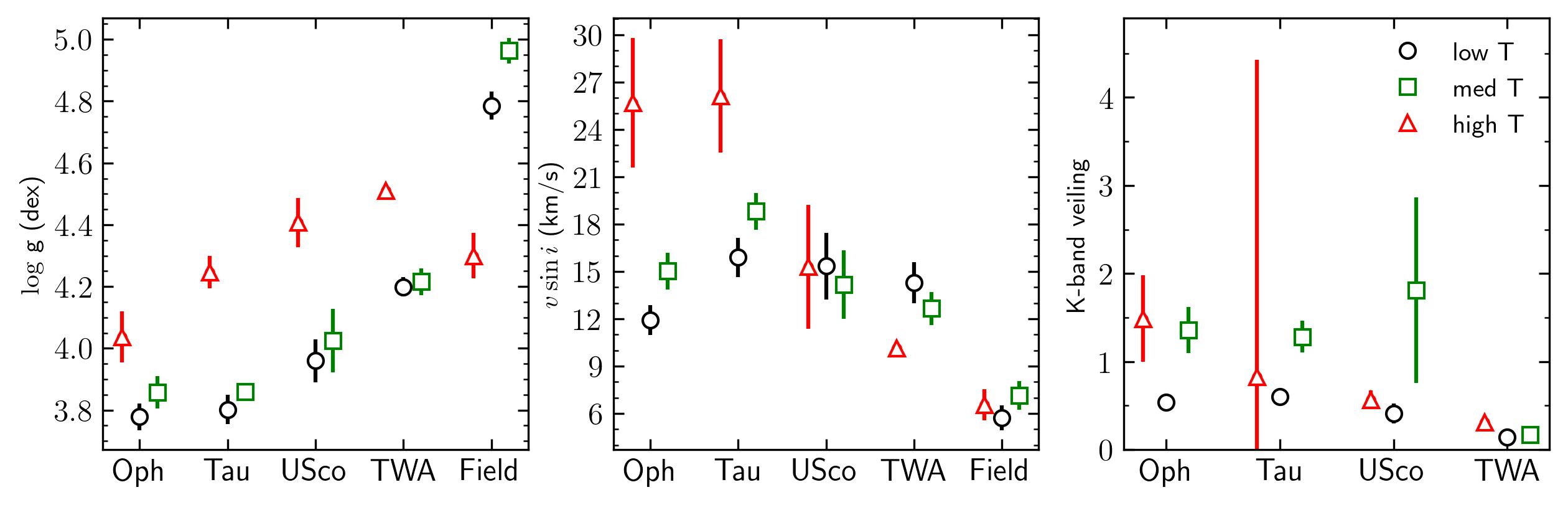}
    \caption{Mean \logg, \vsini, and K-band veiling as a function of \teff\ bins and by star-forming region. The YSOs of Taurus, Ophiuchus, Upper Scorpius, and TWA are divided into low (circles, 3100 $\le$ \teff\ $<$ 3600 K), med (squares, 3600 $\le$ \teff\ $<$ 4100 K), and high (triangles, \teff\ $\geq$ 4100 K) temperature bins. The error bar is the standard error on the mean ($\sigma/\sqrt{n}$) of each \teff\ bin. We also included in the left and middle panels the field stars.}
    \label{fig:mass_bin}
\end{figure*}
}

%------------------------
%-------------------------
%-------------------------
\section{Summary and conclusions}
Using high signal-to-noise K-band spectra obtained with IGRINS and an MCMC approach to spectral fitting, we determined stellar parameters for a sample of {61} YSOs in the Ophiuchus and Upper Scorpius complexes. We also determined the stellar parameters for some K and M field stars reported in \cite{lopezval19}.
We simultaneously fit four observed spectral regions with a synthetic spectral grid computed with solar metallicity MARCS models and the spectral synthesis code {\sc moogstokes} to obtain \teff, \logg, \vsini, \rk\ and B field strength. 
Our B field determination relies on the Zeeman broadening, whose detection is limited by the \vsini\ of the star and the spectral resolution. For this reason, we determined B just for {27} YSOs and provided limits on the B field for the remaining stars in the sample.

We found that our temperatures for M0-M3 stars agree with
three published temperature scales \citep{luhman03b,herczeg14,pecaut13}. For late M stars, our \teff\ values agree more with the \cite{luhman03b} while for K stars, we found that our temperatures are colder than the temperature scales of \cite{pecaut13} and \cite{herczeg14}. 
The differences in the \teff\ can be explained as a cumulative result of small number statistics, differences in the employed methodology, use of different atmospheric models, line lists, and wavelength intervals, the effect of the B field on \teff, and inconsistent spectral type classifications.

{We found that the mean \logg\ value of Ophiuchus (3.83$\pm$0.03 dex) is similar to that of Taurus (3.87$\pm$0.03 dex), while the mean \logg\ of Upper Scorpius (4.13$\pm$0.08 dex) is similar to the value found for TWA (4.22$\pm$0.03 dex). If we consider \logg\ as a proxy for age, our results suggest that the Ophiuchus and Taurus samples have a similar age. We also find that Upper Scorpius and TWA YSOs have similar ages, and are more evolved than Ophiuchus/Taurus YSOs.}
We derived visual extinction and gathered from literature 2MASS and WISE photometry to construct de-reddened infrared SEDs for our sample. We compute spectral indices and classify our YSO samples into morphological classes with these SEDs. 
In the combined sample of YSOs, which includes objects from Ophiuchus, Taurus, Upper Scorpius, and TWA,  we found that Class {\sc ii} YSOs have, on average, lower \logg\ and \vsini, similar B field, and higher K-veiling values than their Class {\sc iii} counterparts.

%--------------------
%--0-------------------
%-----------------------
\begin{acknowledgements}
We appreciate the detailed comments from the anonymous referee, which improved the science and text of this manuscript. This work used the Immersion Grating Infrared Spectrograph (IGRINS) that was developed under a 
collaboration between the University of Texas at Austin and the Korea Astronomy and Space Science 
Institute (KASI) with the financial support of the US National Science Foundation under grant 
AST-1229522 and AST-1702267, of the University of Texas at Austin, and of the Korean GMT Project of 
KASI. RLV acknowledges support from CONACYT through a postdoctoral fellowship within the program ‘Estancias Posdoctorales por México’.

These results made use of the Lowell Discovery Telescope. Lowell is a 
private, non-profit institution dedicated to astrophysical research and public appreciation of 
astronomy and operates the LDT in partnership with Boston University, the University of Maryland, the 
University of Toledo, Northern Arizona University and Yale University. This paper includes data taken 
at The McDonald Observatory of The University of Texas at Austin. Based on observations obtained at the Gemini Observatory, which is operated by the Association of Universities for Research in Astronomy, Inc., under a cooperative agreement with the NSF on behalf of the Gemini partnership: the National Science Foundation (United States), the National Research Council (Canada), CONICYT (Chile), Ministerio de Ciencia, Tecnolog\'{i}a e Innovaci\'{o}n Productiva (Argentina), and Minist\'{e}rio da Ci\^{e}ncia, Tecnologia e Inova\c{c}\~{a}o (Brazil).
This material is based upon work supported by the National Science Foundation under Grant No.\ AST-1908892 to G.\ Mace.
\end{acknowledgements}
\software{IGRINS pipeline package \citep{lee17}, \texttt{MoogStokes} \citep{deen13}, \texttt{MARCS} models \citep{marcs}, \texttt{zbarycorr} \citep{wright2014}, \texttt{MassAge} (Hernandez et al. in prep)}

\appendix

\section{Stellar parameters of Taurus, TWA, and field stars}

\subsection{Taurus and TWA}\label{app:tau}
We included in this work the Taurus and TWA YSOs analyzed in \citetalias{lopezv21}. We compare their stellar parameters to establish similarities and differences between these four YSO populations. \\
In Table~\ref{tab:tau}, we gathered the stellar parameters reported in \citetalias{lopezv21}\ for the Taurus and TWA YSOs. We also added the values of \ext\ and \alfa\ computed in this work for many of them. 

\startlongtable
\begin{deluxetable*}{llllllllllll}
\tabletypesize{\scriptsize}   
\tablewidth{700pt}
 \tablecaption{Stellar parameters for Taurus and TWA YSOs. Columns 1-3 provide target names, spectral types, and membership information. The columns 4 and 5 are the values of the visual extinction and the infrared index c omputed in this work. Finally, column 6-11 present the quality flag, \teff, \logg, B field, \vsini, and K-band veiling determined in \citet{lopezv21}.\label{tab:tau}}
\tablehead{
\colhead{2MASS} & \colhead{SpT} & \colhead{cluster} & \colhead{\ext} & \colhead{$\alpha_{K_s-W4}$} & \colhead{flag} & \colhead{\teff} & \colhead{\logg} & \colhead{B} & \colhead{\vsini} & \colhead{$r_{\rm K}$} \\
& &  & \colhead{mag} & & & \colhead{K} & \colhead{dex} & \colhead{kG} & \colhead{km $s^{-1}$} & }
\startdata
J04043984+2158215 & M3.2 & Tau & 0.7$\pm$0.6 & -2.6$\pm$0.0 & 1 & 3461$\pm$216 & 4.54$\pm$0.35 & 2.75$\pm$0.60 & 17.7$\pm$3.4 & 0.09$\pm$0.06 \\
J04053087+2151106 & M2.7 & Tau & 0.6$\pm$0.6 & -2.6$\pm$0.0 & 1 & 3496$\pm$183 & 4.61$\pm$0.24 & 3.21$\pm$0.39 & 5.0$\pm$2.4  & 0.04$\pm$0.03 \\
J04131414+2819108 & M3.6 & Tau & 0.4$\pm$0.4 & -2.8$\pm$0.1 & 0 & 3393$\pm$154 & 3.40$\pm$0.24 & $<$1.43       & 31.2$\pm$2.3 & 0.23$\pm$0.06 \\
J04132722+2816247 & M0.5 & Tau & 2.6$\pm$0.3 & -2.8$\pm$0.1 & 1 & 3875$\pm$181 & 3.95$\pm$0.34 & $<$2.49       & 27.3$\pm$4.0 & 0.38$\pm$0.11 \\
J04141358+2812492 & M4.5 & Tau & 0.9$\pm$0.5 & -0.6$\pm$0.2 & 0 & 3318$\pm$220 & 4.19$\pm$0.42 & 2.67$\pm$0.60 & 9.7$\pm$3.2  & 1.62$\pm$0.21 \\
J04141458+2827580 & M3.5 & Tau & 0.8$\pm$0.5 & -0.4$\pm$0.2 & 0 & 3251$\pm$157 & 3.34$\pm$0.27 & 1.14$\pm$0.33 & 4.6$\pm$2.3  & 0.37$\pm$0.08 \\
J04141760+2806096 & M5.0 & Tau & 1.9$\pm$0.6 & -0.3$\pm$0.1 & 0 & 3281$\pm$196 & 3.88$\pm$0.40 & $<$1.09       & 9.9$\pm$2.9  & 1.10$\pm$0.21 \\
J04143054+2805147 & M2.2 & Tau & 5.1$\pm$0.6 & 0.2$\pm$0.1  & 0 & 3620$\pm$186 & 3.64$\pm$0.34 & $<$2.36       & 22.7$\pm$3.9 & 0.73$\pm$0.12 \\
J04144730+2646264 & M2.6 & Tau & 0.7$\pm$0.8 & -1.2$\pm$0.2 & 0 & 3530$\pm$257 & 3.72$\pm$0.42 & $<$1.72       & 31.0$\pm$4.2 & 0.44$\pm$0.15 \\
J04144786+2648110 & M2.5 & Tau & 0.5$\pm$0.5 & -0.7$\pm$0.2 & 0 & 3520$\pm$180 & 3.61$\pm$0.30 & $<$1.39       & 20.5$\pm$2.5 & 0.28$\pm$0.08 \\
J04144928+2812305 & M3.5 & Tau & 1.6$\pm$0.3 & -0.9$\pm$0.1 & 0 & 3246$\pm$104 & 3.47$\pm$0.21 & $<$1.03       & 11.1$\pm$1.9 & 0.52$\pm$0.06 \\
J04154278+2909597 & M1.0 & Tau & 2.2$\pm$0.3 & -1.4$\pm$0.5 & 0 & 3723$\pm$125 & 3.87$\pm$0.22 & 1.36$\pm$0.32 & 8.2$\pm$2.0  & 0.12$\pm$0.04 \\
J04162810+2807358 & M2.0 & Tau & 1.1$\pm$0.3 & -2.7$\pm$0.0 & 1 & 3713$\pm$138 & 3.85$\pm$0.21 & $<$2.26       & 27.3$\pm$2.3 & 0.23$\pm$0.04 \\
J04173372+2820468 & M2.3 & Tau & 0.1$\pm$0.5 & -1.3$\pm$0.1 & 0 & 3445$\pm$155 & 3.73$\pm$0.28 & $<$1.31       & 10.8$\pm$2.3 & 0.81$\pm$0.13 \\
J04173893+2833005 & M2.2 & Tau & 0.9$\pm$0.3 & -2.6$\pm$0.0 & 1 & 3658$\pm$136 & 4.06$\pm$0.25 & $<$2.53       & 34.4$\pm$2.6 & 0.15$\pm$0.05 \\
J04174965+2829362 & M3.7 & Tau & 1.4$\pm$0.3 & -0.8$\pm$0.1 & 0 & 3315$\pm$129 & 3.37$\pm$0.25 & 1.22$\pm$0.33 & 9.7$\pm$2.0  & 0.28$\pm$0.06 \\
J04181078+2519574 & M1.0 & Tau &-0.7$\pm$0.3 & -0.7$\pm$0.1 & 0 & 3721$\pm$149 & 3.67$\pm$0.23 & $<$1.12       & 15.6$\pm$2.2 & 0.32$\pm$0.07 \\
J04182909+2826191 & M3.5 & Tau & --          & --           & 1 & 3618$\pm$327 & 4.05$\pm$0.59 & $<$2.32       & 30.7$\pm$7.4 & 0.39$\pm$0.20\\
J04183112+2816290 & M3.5 & Tau & 2.0$\pm$0.6 & -0.5$\pm$0.1 & 0 & 3250$\pm$207 & 3.47$\pm$0.37 & $<$1.48       & 16.9$\pm$3.1 & 1.82$\pm$0.26 \\
J04183158+2816585 & M4.2 & Tau & 1.3$\pm$0.7 & 0.2$\pm$0.2  & 1 & 3385$\pm$210 & 3.83$\pm$0.35 & $<$1.70       & 31.4$\pm$3.1 & 0.64$\pm$0.12 \\
J04183444+2830302 & M0.0 & Tau & --          & --           & 0 & 4230$\pm$142 & 4.53$\pm$0.27 & 1.84$\pm$0.49 & 10.9$\pm$3.0 & 0.44$\pm$0.08 \\
J04184703+2820073 & K8.0 & Tau & 1.8$\pm$0.1 & -2.6$\pm$0.0 & 1 & 3806$\pm$81  & 3.69$\pm$0.14 & 2.63$\pm$0.27 & 16.8$\pm$1.7 & 0.26$\pm$0.01 \\
J04191281+2829330 & M3.5 & Tau & 1.4$\pm$0.4 & -1.2$\pm$0.2 & 1 & 3296$\pm$170 & 3.93$\pm$0.30 & $<$1.43       & 18.9$\pm$2.4 & 0.63$\pm$0.10 \\
J04191583+2906269 & M0.5 & Tau & 0.2$\pm$0.3 & -0.9$\pm$0.1 & 0 & 3719$\pm$130 & 4.01$\pm$0.25 & 2.19$\pm$0.37 & 9.9$\pm$2.3  & 1.36$\pm$0.11 \\
J04192625+2826142 & K8.0 & Tau & 1.1$\pm$0.2 & -2.2$\pm$0.2 & 0 & 3830$\pm$124 & 3.91$\pm$0.21 & 2.22$\pm$0.30 & 9.8$\pm$2.0  & 0.20$\pm$0.03 \\
J04194127+2749484 & M0.0 & Tau & 0.9$\pm$0.2 & -2.7$\pm$0.0 & 0 & 3853$\pm$153 & 4.08$\pm$0.26 & $<$1.94       & 15.6$\pm$2.5 & 0.19$\pm$0.05 \\
J04202606+2804089 & M3.5 & Tau & 0.4$\pm$0.6 & -0.5$\pm$0.3 & 0 & 3407$\pm$197 & 4.09$\pm$0.35 & 1.37$\pm$0.63 & 10.5$\pm$2.8 & 0.22$\pm$0.10 \\
J04214323+1934133 & M2.4 & Tau & 3.5$\pm$0.6 & -0.5$\pm$0.1 & 2 & 3543$\pm$168 & 3.27$\pm$0.27 & 1.04$\pm$0.35 & 7.9$\pm$2.2  & 0.70$\pm$0.09 \\
J04215563+2755060 & M2.3 & Tau & 0.3$\pm$0.5 & -0.9$\pm$0.1 & 0 & 3463$\pm$157 & 3.45$\pm$0.27 & $<$1.01       & 9.4$\pm$2.2  & 1.11$\pm$0.11 \\
J04215943+1932063 & K0.0 & Tau & 1.1$\pm$0.4 & 0.0$\pm$0.1  & 0 & 4065$\pm$232 & 3.51$\pm$0.40 & $<$2.34       & 21.5$\pm$4.2 & 4.90$\pm$0.85 \\
J04220217+2657304 & M0.0 & Tau & 2.6$\pm$0.3 & -0.3$\pm$0.2 & 1 & 3810$\pm$186 & 4.12$\pm$0.36 & $<$2.34       & 20.8$\pm$3.7 & 1.38$\pm$0.19 \\
J04220313+2825389 & M2.5 & Tau & 1.2$\pm$0.4 & -2.4$\pm$0.1 & 0 & 3659$\pm$226 & 4.17$\pm$0.36 & $<$2.67       & 43.7$\pm$3.8 & 0.13$\pm$0.09 \\
J04221675+2654570 & M1.5 & Tau & 3.7$\pm$0.6 & -0.8$\pm$0.1 & 0 & 3586$\pm$236 & 3.83$\pm$0.44 & 2.20$\pm$0.77 & 16.4$\pm$3.5 & 1.74$\pm$0.26 \\
J04244457+2610141 & M2.8 & Tau & 4.1$\pm$0.7 & -0.8$\pm$0.0 & 1 & 3624$\pm$356 & 4.01$\pm$0.64 & 1.55$\pm$0.98 & 11.1$\pm$5.5 & 2.47$\pm$0.63 \\
J04245708+2711565 & M0.6 & Tau & 0.7$\pm$0.4 & -1.0$\pm$0.1 & 0 & 3686$\pm$183 & 4.16$\pm$0.28 & 2.46$\pm$0.43 & 11.5$\pm$2.5 & 1.23$\pm$0.15 \\
J04265352+2606543 & M0.0 & Tau & 4.7$\pm$0.4 & --           & 0 & 4034$\pm$182 & 3.74$\pm$0.35 & $<$2.36       & 37.5$\pm$3.2 & 0.95$\pm$0.20 \\
J04265440+2606510 & M2.5 & Tau & 3.5$\pm$0.5 & -1.1$\pm$0.1 & 0 & 3456$\pm$136 & 3.53$\pm$0.23 & $<$2.01       & 38.6$\pm$2.4 & 0.33$\pm$0.06 \\
J04270469+2606163 & K7.0 & Tau & 2.1$\pm$0.4 & -0.0$\pm$0.1 & 2 & 3969$\pm$216 & 3.20$\pm$0.28 & $<$1.78       & 26.9$\pm$4.3 & 4.82$\pm$0.94 \\
J04293606+2435556 & M3.0 & Tau & 3.1$\pm$0.6 & -1.3$\pm$0.2 & 1 & 3557$\pm$172 & 3.74$\pm$0.28 & $<$2.28       & 24.6$\pm$2.7 & 0.38$\pm$0.08 \\
J04294155+2632582 & M2.3 & Tau & 0.3$\pm$0.3 & -1.0$\pm$0.3 & 0 & 3477$\pm$125 & 3.89$\pm$0.22 & 2.21$\pm$0.32 & 8.4$\pm$2.1  & 1.24$\pm$0.10 \\
J04294247+2632493 & M0.7 & Tau & 0.8$\pm$0.2 & -1.7$\pm$0.4 & 0 & 3675$\pm$102 & 3.61$\pm$0.17 & $<$1.05       & 12.2$\pm$1.8 & 0.23$\pm$0.03 \\
J04295156+2606448 & M1.1 & Tau & 2.6$\pm$0.5 & -1.0$\pm$0.1 & 0 & 3612$\pm$176 & 3.72$\pm$0.31 & 1.81$\pm$0.53 & 14.3$\pm$2.5 & 1.42$\pm$0.15 \\
J04300357+1813494 & M2.0 & Tau & 0.6$\pm$0.4 & -2.7$\pm$0.0 & 0 & 3527$\pm$127 & 4.07$\pm$0.19 & 2.41$\pm$0.32 & 11.7$\pm$2.1 & 0.02$\pm$0.03 \\
J04300399+1813493 & K0.0 & Tau & 1.9$\pm$0.2 & -1.1$\pm$0.4 & 0 & 4606$\pm$233 & 4.43$\pm$0.42 & $<$1.70       & 26.0$\pm$4.5 & 1.81$\pm$0.44 \\
J04302961+2426450 & M2.2 & Tau & 2.2$\pm$0.3 & -1.1$\pm$0.2 & 0 & 3587$\pm$114 & 3.82$\pm$0.21 & $<$1.02       & 8.3$\pm$2.0  & 0.61$\pm$0.07 \\
J04304425+2601244 & K8.5 & Tau & 3.2$\pm$0.3 & -0.9$\pm$0.0 & 1 & 3809$\pm$183 & 4.00$\pm$0.35 & 2.55$\pm$0.64 & 17.6$\pm$3.3 & 3.16$\pm$0.29 \\
J04305137+2442222 & M4.3 & Tau & 0.9$\pm$0.4 & -1.2$\pm$0.1 & 1 & 3277$\pm$144 & 3.61$\pm$0.27 & $<$1.52       & 18.4$\pm$2.4 & 0.41$\pm$0.08 \\
J04311444+2710179 & K8.0 & Tau & 0.3$\pm$0.2 & -2.2$\pm$0.2 & 0 & 3802$\pm$139 & 4.06$\pm$0.26 & 2.01$\pm$0.41 & 13.9$\pm$2.4 & 0.12$\pm$0.05 \\
J04312382+2410529 & M4.5 & Tau & 0.9$\pm$0.6 & -2.6$\pm$0.1 & 1 & 3203$\pm$171 & 4.14$\pm$0.37 & $<$1.55       & 16.6$\pm$3.1 & 0.92$\pm$0.14 \\
J04314007+1813571 & M2.0 & Tau & -0.1$\pm$0.9& -0.3$\pm$0.2 & 2 & 3606$\pm$262 & 3.24$\pm$0.35 & $<$1.87       & 17.1$\pm$3.6 & 3.85$\pm$0.56 \\
J04315056+2424180 & M1.0 & Tau & 2.3$\pm$0.6 & -0.6$\pm$0.3 & 0 & 3598$\pm$203 & 3.64$\pm$0.35 & $<$1.58       & 22.9$\pm$2.9 & 0.93$\pm$0.14 \\
J04315779+1821350 & M3.3 & Tau & 2.6$\pm$0.4 & --           & 1 & 3492$\pm$161 & 3.74$\pm$0.29 & $<$1.84       & 23.0$\pm$2.9 & 0.36$\pm$0.09 \\
J04315779+1821380 & M1.7 & Tau & 2.2$\pm$0.5 & --           & 0 & 3518$\pm$162 & 3.65$\pm$0.26 & $<$1.93       & 18.2$\pm$2.5 & 0.30$\pm$0.07 \\
J04320926+1757227 & K6.0 & Tau & 0.4$\pm$0.1 & -2.7$\pm$0.0 & 0 & 4122$\pm$96  & 4.20$\pm$0.18 & $<$2.62       & 30.9$\pm$2.1 & 0.16$\pm$0.04 \\
J04321456+1820147 & M2.0 & Tau & 0.7$\pm$0.2 & -2.7$\pm$0.0 & 1 & 3610$\pm$91  & 3.91$\pm$0.15 & $<$2.42       & 20.8$\pm$1.8 & 0.13$\pm$0.02 \\
J04321885+2422271 & M0.8 & Tau & 2.0$\pm$0.4 & -2.8$\pm$0.1 & 0 & 3684$\pm$139 & 3.65$\pm$0.22 & $<$1.88       & 32.1$\pm$2.3 & 0.15$\pm$0.05 \\
J04323034+1731406 & K7.5 & Tau & 0.8$\pm$0.3 & -0.8$\pm$0.1 & 0 & 3711$\pm$121 & 3.68$\pm$0.20 & $<$1.10       & 10.4$\pm$1.9 & 0.98$\pm$0.07 \\
J04323058+2419572 & M0.1 & Tau & 3.2$\pm$0.5 & -1.2$\pm$0.2 & 0 & 3925$\pm$222 & 3.90$\pm$0.37 & 1.71$\pm$0.76 & 13.0$\pm$3.1 & 0.78$\pm$0.14 \\
J04323176+2420029 & M0.5 & Tau & 3.0$\pm$0.8 & -0.9$\pm$0.1 & 2 & 3751$\pm$318 & 3.37$\pm$0.48 & 1.20$\pm$0.88 & 8.2$\pm$4.6  & 5.53$\pm$1.05 \\
J04324282+2552314 & M2.0 & Tau & 0.6$\pm$0.5 & --           & 0 & 3426$\pm$179 & 3.79$\pm$0.33 & $<$1.47       & 18.6$\pm$2.5 & 0.24$\pm$0.10 \\
J04324303+2552311 & M1.9 & Tau & 1.6$\pm$1.8 & -0.8$\pm$0.1 & 1 & 3670$\pm$437 & 3.90$\pm$0.64 & $<$2.23       & 31.0$\pm$9.7 & 3.06$\pm$0.97 \\
J04324373+1802563 & K6.0 & Tau & 0.4$\pm$0.2 & -2.7$\pm$0.1 & 0 & 3912$\pm$126 & 4.11$\pm$0.23 & 1.40$\pm$0.35 & 7.5$\pm$2.2  & 0.12$\pm$0.04 \\
J04324911+2253027 & K5.5 & Tau & 3.6$\pm$0.6 & -0.8$\pm$0.2 & 0 & 4105$\pm$394 & 4.15$\pm$0.64 & $<$2.40       & 38.2$\pm$8.2 & 1.58$\pm$0.59 \\
J04324938+2253082 & M4.5 & Tau & 3.9$\pm$0.9 & -1.2$\pm$0.1 & 2 & 3380$\pm$247 & 3.42$\pm$0.42 & $<$1.18       & 18.1$\pm$3.4 & 0.37$\pm$0.14 \\
J04325323+1735337 & M2.0 & Tau & 0.5$\pm$0.5 & -2.0$\pm$0.2 & 0 & 3543$\pm$132 & 3.65$\pm$0.22 & $<$1.10       & 11.4$\pm$2.0 & 0.09$\pm$0.04 \\
J04330622+2409339 & M2.0 & Tau & 1.0$\pm$0.5 & -1.1$\pm$0.2 & 0 & 3553$\pm$160 & 3.64$\pm$0.26 & $<$1.59       & 28.1$\pm$2.5 & 0.57$\pm$0.09 \\
J04330664+2409549 & K7.0 & Tau & 1.0$\pm$0.3 & -1.4$\pm$0.2 & 1 & 3880$\pm$177 & 4.20$\pm$0.35 & 2.50$\pm$0.71 & 19.9$\pm$4.5 & 1.16$\pm$0.15 \\
J04331003+2433433 & K7.5 & Tau & 0.5$\pm$0.1 & -2.8$\pm$0.1 & 0 & 3878$\pm$80  & 3.89$\pm$0.14 & $<$2.40       & 31.6$\pm$1.8 & 0.16$\pm$0.02 \\
J04333405+2421170 & M0.4 & Tau & 1.9$\pm$0.4 & -0.7$\pm$0.1 & 0 & 3689$\pm$185 & 3.79$\pm$0.32 & 1.92$\pm$0.44 & 12.1$\pm$2.6 & 1.42$\pm$0.15 \\
J04333456+2421058 & K6.5 & Tau & 1.5$\pm$0.3 & -0.7$\pm$0.1 & 0 & 3842$\pm$203 & 3.83$\pm$0.33 & $<$2.05       & 21.2$\pm$3.0 & 1.57$\pm$0.18 \\
J04333678+2609492 & M0.0 & Tau & 1.9$\pm$0.4 & -0.9$\pm$0.1 & 0 & 3581$\pm$152 & 3.98$\pm$0.25 & 2.28$\pm$0.37 & 11.8$\pm$2.4 & 0.71$\pm$0.09 \\
J04333906+2520382 & K5.5 & Tau & 0.9$\pm$0.5 & -0.8$\pm$0.1 & 0 & 3930$\pm$222 & 3.80$\pm$0.43 & 1.95$\pm$0.70 & 11.0$\pm$3.7 & 3.23$\pm$0.42 \\
J04334871+1810099 & M3.0 & Tau & 0.1$\pm$0.3 & -0.6$\pm$0.5 & 0 & 3449$\pm$103 & 4.09$\pm$0.18 & 1.63$\pm$0.29 & 5.7$\pm$2.0  & 0.08$\pm$0.03 \\
J04335200+2250301 & K5.5 & Tau & 1.3$\pm$0.2 & -0.9$\pm$0.1 & 0 & 3951$\pm$94  & 3.77$\pm$0.17 & 1.95$\pm$0.31 & 12.5$\pm$1.9 & 2.34$\pm$0.09 \\
J04335470+2613275 & K5.0 & Tau & 2.6$\pm$0.4 & -1.2$\pm$0.1 & 0 & 3977$\pm$195 & 4.09$\pm$0.35 & $<$2.83       & 36.4$\pm$3.8 & 0.71$\pm$0.14 \\
J04341099+2251445 & M1.5 & Tau & 1.8$\pm$0.5 & -2.7$\pm$0.0 & 0 & 3589$\pm$154 & 3.97$\pm$0.26 & 2.16$\pm$0.40 & 13.3$\pm$2.3 & 0.09$\pm$0.06 \\
J04345542+2428531 & M0.6 & Tau & 4.9$\pm$0.3 & -0.8$\pm$0.2 & 0 & 3751$\pm$171 & 3.87$\pm$0.26 & 2.16$\pm$0.41 & 12.5$\pm$2.3 & 1.56$\pm$0.12 \\
J04352020+2232146 & M3.2 & Tau & 0.5$\pm$0.7 & -1.1$\pm$0.1 & 0 & 3493$\pm$240 & 4.00$\pm$0.48 & $<$1.62       & 20.2$\pm$3.8 & 1.04$\pm$0.22 \\
J04352089+2254242 & K8.0 & Tau & 1.9$\pm$0.4 & -2.8$\pm$0.1 & 0 & 3856$\pm$195 & 3.87$\pm$0.30 & $<$0.90       & 6.2$\pm$2.4  & 0.17$\pm$0.06 \\
J04352450+1751429 & M2.6 & Tau & 0.9$\pm$0.4 & -2.6$\pm$0.0 & 1 & 3494$\pm$146 & 4.28$\pm$0.21 & 2.44$\pm$0.33 & 5.0$\pm$2.3  & 0.04$\pm$0.04 \\
J04352737+2414589 & M0.3 & Tau & 0.1$\pm$0.4 & -0.9$\pm$0.2 & 0 & 3648$\pm$144 & 3.74$\pm$0.22 & 1.54$\pm$0.37 & 10.9$\pm$2.1 & 0.43$\pm$0.05 \\
J04354093+2411087 & M0.5 & Tau & 4.6$\pm$0.4 & -1.0$\pm$0.0 & 0 & 3665$\pm$154 & 3.85$\pm$0.26 & 2.33$\pm$0.34 & 8.2$\pm$2.1  & 0.77$\pm$0.08 \\
J04354733+2250216 & K2.0 & Tau & 2.4$\pm$0.3 & -0.7$\pm$0.1 & 0 & 4265$\pm$255 & 4.22$\pm$0.44 & $<$2.85       & 43.0$\pm$5.0 & 1.18$\pm$0.32 \\
J04355277+2254231 & K4.0 & Tau & 3.0$\pm$0.5 & -0.6$\pm$0.2 & 1 & 4096$\pm$380 & 4.07$\pm$0.62 & $<$1.79       & 18.6$\pm$8.8 & 2.23$\pm$0.90 \\
J04355349+2254089 & M0.6 & Tau & 2.3$\pm$0.3 & --           & 0 & 3760$\pm$164 & 3.84$\pm$0.26 & $<$2.65       & 43.3$\pm$2.9 & 0.25$\pm$0.07 \\
J04355684+2254360 & M2.0 & Tau & 2.0$\pm$0.6 & -0.9$\pm$0.1 & 0 & 3406$\pm$207 & 3.99$\pm$0.43 & $<$1.22       & 12.5$\pm$3.1 & 0.88$\pm$0.20 \\
J04355892+2238353 & K8.0 & Tau & 0.9$\pm$0.3 & -2.7$\pm$0.1 & 0 & 3869$\pm$169 & 4.10$\pm$0.29 & $<$2.54       & 29.1$\pm$3.5 & 0.16$\pm$0.07 \\
J04361909+2542589 & K5.0 & Tau & 0.3$\pm$0.2 & -2.9$\pm$0.1 & 0 & 4134$\pm$126 & 4.07$\pm$0.24 & $<$1.72       & 22.8$\pm$2.5 & 0.15$\pm$0.07 \\
J04382858+2610494 & M0.3 & Tau & 1.3$\pm$0.6 & -0.5$\pm$0.1 & 2 & 3704$\pm$350 & 3.44$\pm$0.50 & 1.78$\pm$0.85 & 12.7$\pm$5.4 & 5.44$\pm$1.04 \\
J04390163+2336029 & M4.9 & Tau & 0.1$\pm$0.5 & -1.3$\pm$0.1 & 1 & 3221$\pm$188 & 3.86$\pm$0.36 & $<$1.27       & 17.8$\pm$2.7 & 0.52$\pm$0.12 \\
J04391779+2221034 & K5.5 & Tau & 0.8$\pm$0.2 & -1.1$\pm$0.2 & 0 & 4156$\pm$123 & 4.11$\pm$0.23 & $<$1.83       & 15.4$\pm$2.3 & 1.08$\pm$0.10 \\
J04392090+2545021 & M2.5 & Tau & 3.9$\pm$0.4 & -1.0$\pm$0.1 & 0 & 3536$\pm$152 & 3.76$\pm$0.28 & $<$1.58       & 13.8$\pm$2.4 & 1.47$\pm$0.15 \\
J04400800+2605253 & ~    & Tau & 7.2$\pm$0.5 & -0.6$\pm$0.1 & 0 & 3749$\pm$273 & 3.84$\pm$0.49 & 2.32$\pm$0.97 & 16.6$\pm$4.3 & 1.52$\pm$0.28 \\
J04410470+2451062 & M0.9 & Tau & 0.7$\pm$0.2 & -2.7$\pm$0.0 & 0 & 3663$\pm$115 & 4.07$\pm$0.19 & 2.28$\pm$0.29 & 8.3$\pm$2.0  & 0.10$\pm$0.04 \\
J04411681+2840000 & M1.1 & Tau & 2.4$\pm$0.3 & -0.6$\pm$0.6 & 0 & 3706$\pm$121 & 3.70$\pm$0.21 & $<$1.90       & 26.3$\pm$2.2 & 0.22$\pm$0.04 \\
J04413882+2556267 & M0.0 & Tau & 3.3$\pm$0.3 & -0.1$\pm$0.1 & 0 & 3738$\pm$161 & 3.74$\pm$0.26 & $<$1.80       & 24.2$\pm$2.6 & 0.59$\pm$0.09 \\
J04420548+2522562 & M0.5 & Tau & 3.0$\pm$0.3 & -2.2$\pm$0.1 & 0 & 3779$\pm$146 & 3.89$\pm$0.24 & $<$2.22       & 21.7$\pm$2.6 & 0.25$\pm$0.05 \\
J04420777+2523118 & K7.0 & Tau & 3.0$\pm$0.3 & --           & 0 & 3763$\pm$160 & 3.93$\pm$0.28 & 1.59$\pm$0.49 & 11.0$\pm$2.5 & 1.40$\pm$0.12 \\
J04423769+2515374 & M0.8 & Tau & 0.7$\pm$0.5 & -0.3$\pm$0.1 & 0 & 3758$\pm$282 & 3.91$\pm$0.43 & $<$1.89       & 19.9$\pm$4.1 & 1.60$\pm$0.25 \\
J04430309+2520187 & M2.3 & Tau & 1.0$\pm$0.6 & -0.9$\pm$0.2 & 0 & 3487$\pm$181 & 3.89$\pm$0.31 & $<$1.30       & 12.3$\pm$2.5 & 0.49$\pm$0.11 \\
J04465305+1700001 & M0.6 & Tau & 1.3$\pm$0.8 & -0.7$\pm$0.1 & 1 & 3704$\pm$319 & 3.88$\pm$0.57 & $<$1.86       & 17.7$\pm$4.9 & 0.59$\pm$0.23 \\
J04465897+1702381 & K8.0 & Tau & 2.4$\pm$0.4 & -1.1$\pm$0.1 & 0 & 3903$\pm$243 & 3.53$\pm$0.35 & $<$1.30       & 12.8$\pm$3.1 & 2.36$\pm$0.36 \\
J04474859+2925112 & M0.4 & Tau & 0.9$\pm$0.2 & -1.2$\pm$0.1 & 0 & 3879$\pm$138 & 3.89$\pm$0.25 & 2.09$\pm$0.41 & 13.4$\pm$2.3 & 1.79$\pm$0.13 \\
J04514737+3047134 & K7.0 & Tau & 0.8$\pm$0.6 & -0.2$\pm$0.0 & 0 & 3780$\pm$279 & 3.79$\pm$0.49 & $<$1.80       & 21.4$\pm$4.3 & 3.03$\pm$0.45 \\
J04551098+3021595 & K6.0 & Tau & 0.9$\pm$0.3 & -0.6$\pm$0.5 & 0 & 4020$\pm$99  & 4.02$\pm$0.19 & 2.11$\pm$0.34 & 14.4$\pm$2.0 & 0.48$\pm$0.05 \\
J04553695+3017553 & K2.0 & Tau & -0.1$\pm$0.1& -2.5$\pm$0.1 & 0 & 4450$\pm$122 & 4.26$\pm$0.25 & $<$1.95       & 21.8$\pm$2.8 & 0.24$\pm$0.09 \\
J04560201+3021037 & K6.0 & Tau & 0.6$\pm$0.3 & -2.4$\pm$0.1 & 0 & 4034$\pm$100 & 3.99$\pm$0.19 & $<$1.61       & 14.2$\pm$2.0 & 0.18$\pm$0.04 \\
J05030659+2523197 & M0.8 & Tau & 1.1$\pm$0.2 & -1.0$\pm$0.2 & 1 & 3760$\pm$158 & 4.01$\pm$0.26 & 2.67$\pm$0.40 & 11.8$\pm$2.4 & 0.68$\pm$0.08 \\
J05071206+2437163 & K6.0 & Tau & 0.5$\pm$0.3 & -2.6$\pm$0.0 & 0 & 3987$\pm$131 & 4.11$\pm$0.25 & $<$2.00       & 21.6$\pm$2.5 & 0.09$\pm$0.05 \\
J05074953+3024050 & K2.0 & Tau & 3.4$\pm$0.7 & -0.8$\pm$0.1 & 0 & 3885$\pm$215 & 4.10$\pm$0.39 & $<$2.02       & 16.6$\pm$3.9 & 2.09$\pm$0.29 \\
J10120908-3124451 & M4.0 & TWA & 1.1$\pm$0.2 & -1.6$\pm$0.1 & 1 & 3315$\pm$87  & 4.06$\pm$0.18 & $<$1.34       & 18.0$\pm$2.0 & 0.28$\pm$0.03 \\
J10423011-3340162 & M3.2 & TWA & --          & --           & 0 & 3327$\pm$80  & 4.27$\pm$0.13 & 2.19$\pm$0.26 & 7.4$\pm$1.7  & 0.10$\pm$0.00 \\
J11015191-3442170 & M0.5 & TWA & --          & --           & 0 & 3783$\pm$107 & 4.35$\pm$0.18 & 2.75$\pm$0.30 & 8.4$\pm$2.0  & 0.53$\pm$0.05 \\
J11091380-3001398 & M2.2 & TWA & --          & --           & 0 & 3557$\pm$92  & 4.07$\pm$0.16 & $<$1.38       & 15.9$\pm$1.9 & 0.06$\pm$0.02\\
J11102788-3731520 & M4.1 & TWA & --          & --           & 1 & 3284$\pm$75  & 4.00$\pm$0.13 & $<$0.97       & 11.9$\pm$1.7 & 0.29$\pm$0.00\\
J11210549-3845163 & M2.75& TWA & --          & --           & 0 & 3534$\pm$104 & 4.19$\pm$0.17 & 2.67$\pm$0.32 & 19.7$\pm$2.0 & 0.02$\pm$0.02 \\
J11211723-3446454 & M1.1 & TWA & --          & --           & 0 & 3638$\pm$95  & 4.14$\pm$0.16 & 2.12$\pm$0.29 & 14.2$\pm$1.9 & 0.04$\pm$0.02 \\
J11211745-3446497 & M1.0 & TWA & --          & --           & 0 & 3672$\pm$93  & 4.13$\pm$0.16 & $<$1.71       & 13.7$\pm$1.9 & 0.04$\pm$0.02\\
J11220530-2446393 & K6.0 & TWA & --          & --           & 0 & 4257$\pm$102 & 4.51$\pm$0.21 & $<$1.10       & 10.2$\pm$2.1 & 0.31$\pm$0.04\\
J11324124-2651559 & M2.9 & TWA & --          & --           & 0 & 3397$\pm$75  & 4.30$\pm$0.13 & 2.86$\pm$0.26 & 7.5$\pm$1.7  & 0.13$\pm$0.00 \\
J11482373-3728485 & M3.4 & TWA & 0.2$\pm$0.2 & -2.7$\pm$0.1 & 0 & 3351$\pm$75  & 4.30$\pm$0.14 & $<$1.30       & 10.9$\pm$1.7 & 0.09$\pm$0.00 \\
J11482422-3728491 & K6.0 & TWA & 0.1$\pm$0.3 & -2.7$\pm$0.0 & 0 & 4043$\pm$94  & 4.32$\pm$0.17 & 2.29$\pm$0.30 & 12.0$\pm$2.0 & 0.17$\pm$0.03 \\
J12072738-3247002 & M3.5 & TWA & --          & --           & 1 & 3404$\pm$101 & 4.15$\pm$0.18 & $<$1.42       & 19.1$\pm$1.9 & 0.16$\pm$0.03\\
J12153072-3948426 & M0.5 & TWA & --          & --           & 0 & 3707$\pm$99  & 4.14$\pm$0.17 & 2.21$\pm$0.29 & 15.0$\pm$1.9 & 0.06$\pm$0.03 \\
J12345629-4538075 & M3.0 & TWA & 0.0$\pm$0.4 & -2.7$\pm$0.0 & 0 & 3445$\pm$90  & 4.15$\pm$0.16 & 1.85$\pm$0.28 & 12.4$\pm$1.9 & 0.07$\pm$0.02 \\
J12350424-4136385 & M3.0 & TWA & --          & --           & 1 & 3357$\pm$75  & 4.23$\pm$0.13 & 2.54$\pm$0.26 & 9.5$\pm$1.7  & 0.10$\pm$0.00 \\
J12354893-3950245 & M4.5 & TWA & 0.9$\pm$0.4 & -2.5$\pm$0.0 & 1 & 3391$\pm$122 & 4.45$\pm$0.22 & $<$1.98       & 22.8$\pm$2.3 & 0.36$\pm$0.04 \\
J12360055-3952156 & M2.5 & TWA & --          & --           & 0 & 3531$\pm$115 & 4.22$\pm$0.18 & 2.23$\pm$0.34 & 14.8$\pm$2.1 & 0.02$\pm$0.02 \\
TWA 3B            & M4.0 & TWA & --          & --           & 1 & 3355$\pm$75  & 4.20$\pm$0.13 & $<$1.78       & 15.8$\pm$1.7 & 0.16$\pm$0.00 \\
\enddata
\end{deluxetable*}

%=--------------------------
\subsection{Field stars}\label{app:field}
Between 2014 and 2018, as part of various scientific projects, IGRINS observed 254 K and M field stars at the McDonald Observatory 2.7 m telescope, Lowell Discovery Telescope, and Gemini South Telescope.
\cite{lopezval19} presented an accurate determination of their temperatures using absorption line-depths of Fe I, OH, and Al I, located at the H-band and BT-Settl models \citep{btsettl}. 

These K and M stars are an excellent sample of more evolved counterparts for our YSO sample. The field stars allow us to complete the age ladder we are creating with the IGRINS YSO survey, establish the differences or similarities between the populations at different evolutionary stages, and identify the role of the stellar parameters at every step.

We started selecting from \cite{lopezval19} just the stars with spectral type between K0 and M5 and with a minimum mean signal-to-noise ratio of $\sim$50 in the K-band, finishing with a sample of {207} field stars. 
Then, we determined the stellar parameters in the field stars following the MCMC approach as in our YSO sample, but with minor variations.

The main differences between the MCMC applied to the field stars and the YSOs are: i) due to the more evolved status of the field stars, there is no need to include any K-band veiling contribution, so we removed this free parameter from the MCMC analysis. ii) We extended the synthetic spectral grid to include \logg\ values up to 5.5 dex, as M field stars might have \logg\ values close to or even higher than 5.0 dex \citep[e.g.,][]{segransan03}. The MARCS library has models with \logg\ equal to 5.5 dex for \teff\ lower than 3900 K. We extended the synthetic grid for \teff\ colder than 3900 K. 
 
As with the YSOs, we classified the stellar parameter determination of the field stars with the same quality flag as the YSOs. We assigned a flag equal to 0, 1, 2, or 3 if the determination is good, acceptable, at the edge of our grid, or poor. We found that almost {36\%} of our field stars (74 of 207) have poor determinations. The reason for this high number is not clear. We identified some common issues in the stellar spectra that prevented our MCMC analysis from converging into a good fit. 
We identify some stars with very high \vsini, narrow atomic or molecular lines, and some double lines. 
The remaining {133} stars, reported in Table \ref{tab:field}, are considered our field sample, and we used them throughout the paper.  

{In Figure \ref{fig:campo}, we compared the \teff\ values determined in here with those values reported by \cite{lopezval19}, color-coded by our \logg\ values. { There is an acceptable agreement between the two studies, but we found a trend with our \logg values. The line-depth temperatures are hotter than our values for objects with \logg$\lesssim$4.8 dex, while the opposite behavior occurs if the \logg\ that we determined is greater than 4.8 dex. {Such trend might be the result of two main things: i) \cite{lopezval19} used a different set of synthetic spectra (BT-Settl models), and ii) they fixed the \logg\ value to 4.5 dex in their determination method for all the sample.}
Although this, the majority of our determinations (97 of 133) are within $\pm$300 K, however, the differences could be as high as 600 K, especially in those objects that have \logg\ values significantly different from the fixed 4.5 dex used in \cite{lopezval19}. 
We also see that the line-depth temperatures piled up about $\sim$4400 K, where the method loses its effectiveness in determining temperature.

\begin{figure}
    \centering
    \includegraphics[width=\columnwidth]{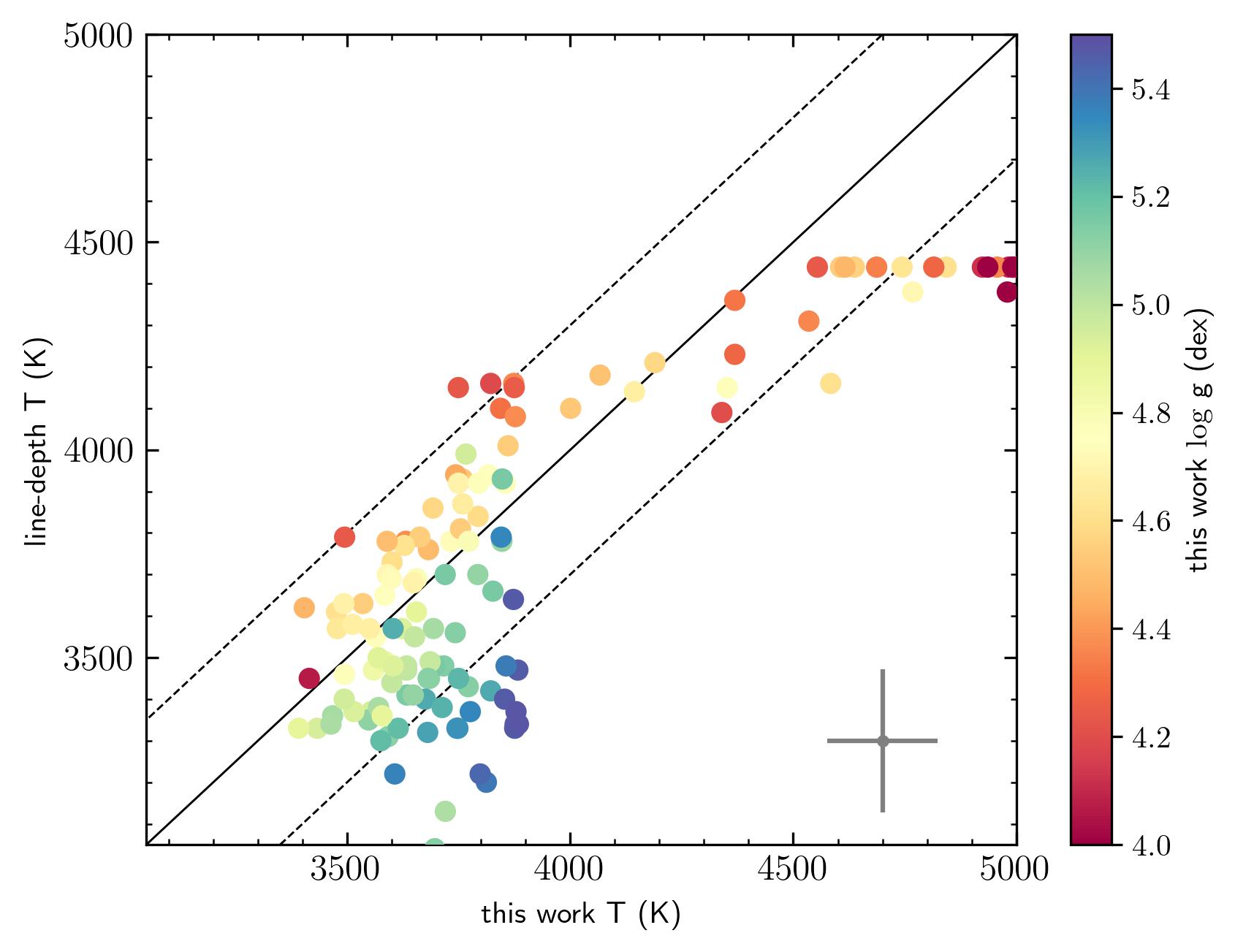}
    \caption{Comparison between our \teff\ values and the line-depth temperatures of \citet{lopezval19} for 133 field stars, color-coded by our \logg\ determinations. The dashed lines are $\pm$300 K from the one-to-one relation (solid line). About 73\% of the sample is within $\pm$300 K, but the differences could be as high as 600 K. The error bar in the right corner represents the mean errors of each method, which are of $\sim$120 and $\sim$170 K, for this work and the line-depth methods, respectively.}
    \label{fig:campo}
\end{figure}

It is important to mention that we selected and fine-tuned the method, synthetic spectra, line list, and the spectral regions used in this paper to determine the stellar parameters of young stellar objects, which may not work properly for more evolved stars. 
Despite this, our stellar parameters follow the theoretical predictions for K and M stars.}

\startlongtable
\begin{deluxetable*}{lllllll}
\tabletypesize{\scriptsize}   
\tablewidth{0pt}
 \tablecaption{Stellar parameters of the K and M field stars. The stellar parameters are determined using the same MCMC approach, but ignoring K-band veiling. The spectral type references are those indicated in \citet{lopezval19}. \label{tab:field}}
 
\tablecolumns{7}
\tablehead{
\colhead{Name} & \colhead{SpT} & \colhead{flag} & \colhead{\teff} & \colhead{\logg} & \colhead{B} & \colhead{\vsini}}
\startdata
[RSP2011] 315 & -- & 0 & 3558$\pm$186 & 4.84$\pm$0.31 & $<$0.42 & 4.7$\pm$2.9 \\
G 43-43 & -- & 0 & 3655$\pm$144 & 4.90$\pm$0.27 & $<$0.32 & 4.0$\pm$2.5 \\
LP 611-70 & -- & 0 & 3494$\pm$149 & 4.24$\pm$0.20 & $<$0.42 & 3.4$\pm$2.0 \\
UCAC4 545-148763 & -- & 0 & 3793$\pm$140 & 4.59$\pm$0.22 & 1.45$\pm$0.39 & 5.5$\pm$2.7 \\
G 194-18 & -- & 2 & 3748$\pm$126 & 5.32$\pm$0.25 & $<$0.40 & 4.9$\pm$3.5 \\
HD 285482 & K0.0 & 0 & 4744$\pm$91 & 4.63$\pm$0.17 & $<$0.34 & 3.9$\pm$2.0 \\
HD 285690 & K0.0 & 0 & 4843$\pm$96 & 4.61$\pm$0.19 & $<$0.33 & 4.3$\pm$2.2 \\
HD 285876 & K0.0 & 0 & 4190$\pm$111 & 4.58$\pm$0.19 & $<$0.26 & 3.4$\pm$2.1 \\
HD 286363 & K0.0 & 0 & 4606$\pm$93 & 4.54$\pm$0.20 & $<$0.36 & 4.5$\pm$2.1 \\
BD+45 598 & K0.0 & 2 & 4924$\pm$111 & 4.12$\pm$0.37 & $<$0.63 & 22.5$\pm$3.6 \\
HD 182488 & K0.0 & 2 & 4983$\pm$78 & 4.09$\pm$0.22 & $<$0.24 & 4.6$\pm$2.1 \\
* 54 Psc & K0.5 & 2 & 4993$\pm$75 & 4.27$\pm$0.15 & $<$0.09 & 3.4$\pm$1.8 \\
* 107 Psc & K1.0 & 2 & 4996$\pm$75 & 4.25$\pm$0.15 & $<$0.05 & 5.0$\pm$1.8 \\
HD 125455 & K1.0 & 2 & 4990$\pm$76 & 4.10$\pm$0.18 & $<$0.13 & 4.8$\pm$1.9 \\
HD 285348 & K2.0 & 0 & 4637$\pm$90 & 4.56$\pm$0.18 & $<$0.37 & 3.3$\pm$2.1 \\
HD 3765 & K2.0 & 0 & 4815$\pm$80 & 4.28$\pm$0.15 & $<$0.09 & 2.3$\pm$1.7 \\
BD+20 2720 & K2.0 & 2 & 4980$\pm$79 & 3.31$\pm$0.24 & $<$0.15 & 3.9$\pm$2.0 \\
HD 21845 & K2.0 & 2 & 4957$\pm$99 & 4.36$\pm$0.28 & $<$0.52 & 18.4$\pm$2.9 \\
HD 220339 & K2.0 & 2 & 4936$\pm$128 & 3.98$\pm$0.46 & $<$0.59 & 8.4$\pm$4.4 \\
HD 88925 & K2.0 & 2 & 4992$\pm$75 & 3.14$\pm$0.19 & $<$0.06 & 5.4$\pm$1.8 \\
HD 122064 & K3.0 & 0 & 4554$\pm$77 & 4.24$\pm$0.14 & $<$0.08 & 2.5$\pm$1.7 \\
HD 219134 & K3.0 & 0 & 4687$\pm$90 & 4.34$\pm$0.18 & $<$0.18 & 3.0$\pm$1.9 \\
HD 52919 & K4.0 & 2 & 4616$\pm$91 & 4.48$\pm$0.21 & $<$0.21 & 2.9$\pm$2.0 \\
BD+49 2125 & K4/5 & 1 & 4535$\pm$130 & 4.36$\pm$0.30 & $<$0.82 & 12.0$\pm$2.6 \\
BD-09 2926B & K5 & 1 & 4584$\pm$106 & 4.61$\pm$0.25 & $<$0.28 & 4.0$\pm$2.3 \\
BD+20 1790 & K5.0 & 0 & 4352$\pm$335 & 4.77$\pm$0.22 & 2.61$\pm$0.49 & 10.8$\pm$2.3 \\
HD 286554 & K5.0 & 0 & 4067$\pm$117 & 4.51$\pm$0.18 & $<$0.28 & 3.3$\pm$2.0 \\
BD-09 4191 & K5.0 & 1 & 3719$\pm$150 & 5.16$\pm$0.30 & $<$0.39 & 7.8$\pm$3.6 \\
HD 98800 & K5.0 & 1 & 4340$\pm$90 & 4.20$\pm$0.18 & $<$0.59 & 10.3$\pm$2.0 \\
HD 122120 & K5.0 & 2 & 4369$\pm$86 & 4.32$\pm$0.16 & $<$0.18 & 2.6$\pm$1.8 \\
BD+05 378 & K6.0 & 0 & 4144$\pm$290 & 4.67$\pm$0.41 & 2.27$\pm$0.72 & 12.1$\pm$3.3 \\
HD 88230 & K6.0 & 0 & 3843$\pm$104 & 4.31$\pm$0.16 & $<$0.26 & 3.2$\pm$2.0 \\
HD 283869 & K7.0 & 0 & 4369$\pm$83 & 4.28$\pm$0.15 & $<$0.15 & 2.7$\pm$1.8 \\
HD 6440B & K7.0 & 0 & 3854$\pm$97 & 4.77$\pm$0.16 & $<$0.17 & 3.3$\pm$1.9 \\
LP 533-57 & K7.0 & 0 & 3873$\pm$83 & 4.36$\pm$0.15 & $<$0.29 & 4.5$\pm$1.9 \\
HD 21845B & K7.0 & 1 & 3758$\pm$198 & 4.53$\pm$0.35 & $<$2.72 & 26.8$\pm$4.6 \\
HD 157881 & K7.0 & 2 & 3877$\pm$80 & 4.37$\pm$0.14 & $<$0.29 & 2.5$\pm$1.8 \\
HD 97101 & K7.0 & 2 & 3874$\pm$80 & 4.25$\pm$0.14 & $<$0.13 & 2.5$\pm$1.8 \\
BD+06 2986 & K8.0 & 1 & 3846$\pm$88 & 5.09$\pm$0.16 & $<$0.12 & 3.1$\pm$1.9 \\
BD-13 6424 & M0.0 & 0 & 3682$\pm$176 & 4.49$\pm$0.26 & 1.96$\pm$0.41 & 9.1$\pm$2.8 \\
BD+02 1729 & M0.0 & 0 & 3743$\pm$91 & 4.44$\pm$0.15 & $<$0.28 & 3.6$\pm$1.9 \\
BD+13 2618 & M0.0 & 0 & 3631$\pm$102 & 4.37$\pm$0.16 & 2.21$\pm$0.29 & 12.5$\pm$1.9 \\
BD+23 2063B & M0.0 & 0 & 3749$\pm$142 & 4.23$\pm$0.21 & $<$0.43 & 3.7$\pm$2.2 \\
BD+33 1505 & M0.0 & 0 & 3692$\pm$117 & 4.58$\pm$0.18 & $<$0.37 & 3.5$\pm$2.0 \\
BD+36 2322 & M0.0 & 0 & 3732$\pm$123 & 4.73$\pm$0.21 & 2.22$\pm$0.36 & 12.8$\pm$2.4 \\
HD 19305 & M0.0 & 0 & 3860$\pm$80 & 4.54$\pm$0.14 & $<$0.15 & 2.6$\pm$1.8 \\
HD 79211 & M0.0 & 0 & 3759$\pm$128 & 4.66$\pm$0.19 & $<$0.23 & 3.1$\pm$2.0 \\
HIP 17248 & M0.0 & 0 & 3754$\pm$194 & 4.54$\pm$0.25 & 2.24$\pm$0.43 & 5.4$\pm$2.8 \\
Ross 1050 & M0.0 & 0 & 3793$\pm$125 & 5.10$\pm$0.24 & $<$0.28 & 4.7$\pm$2.7 \\
StKM 1-82 & M0.0 & 0 & 3816$\pm$126 & 4.76$\pm$0.21 & $<$0.49 & 5.1$\pm$2.4 \\
TYC 5174-242-1 & M0.0 & 0 & 3749$\pm$137 & 4.69$\pm$0.24 & $<$0.68 & 9.9$\pm$2.6 \\
BD+30 397B & M0.0 & 1 & 3563$\pm$147 & 4.74$\pm$0.24 & 2.84$\pm$0.37 & 8.1$\pm$2.3 \\
CCDM J10095+4126AB & M0.0 & 1 & 4768$\pm$115 & 4.71$\pm$0.24 & $<$0.89 & 9.6$\pm$2.6 \\
GSC 04686-00596 & M0.0 & 1 & 4001$\pm$210 & 4.53$\pm$0.31 & $<$2.87 & 37.9$\pm$3.8 \\
HD 59438C & M0.0 & 1 & 3821$\pm$120 & 5.26$\pm$0.26 & $<$0.26 & 10.8$\pm$3.1 \\
StKM 1-516 & M0.0 & 1 & 3826$\pm$102 & 5.16$\pm$0.18 & $<$0.13 & 3.6$\pm$2.1 \\
HD 209290 & M0.5 & 0 & 3662$\pm$100 & 4.55$\pm$0.16 & $<$0.32 & 2.8$\pm$1.9 \\
HD 28343 & M0.5 & 0 & 3821$\pm$96 & 4.19$\pm$0.15 & $<$0.31 & 3.0$\pm$1.9 \\
GJ 3653 & M0.5 & 1 & 3794$\pm$145 & 4.77$\pm$0.29 & $<$2.18 & 34.6$\pm$3.8 \\
MCC 124 & M0.7 & 1 & 3847$\pm$104 & 5.16$\pm$0.20 & 1.60$\pm$0.53 & 9.2$\pm$3.0 \\
2MASS J14040922+2044314 & M1.0 & 0 & 3655$\pm$141 & 4.78$\pm$0.24 & $<$0.78 & 5.3$\pm$2.5 \\
BD+61 1678C & M1.0 & 0 & 3627$\pm$123 & 4.62$\pm$0.19 & $<$0.34 & 3.6$\pm$2.1 \\
HD 199305 & M1.0 & 0 & 3600$\pm$88 & 4.60$\pm$0.15 & $<$0.21 & 2.5$\pm$1.8 \\
HD 42581 & M1.0 & 0 & 3589$\pm$147 & 4.50$\pm$0.21 & $<$0.40 & 3.5$\pm$2.2 \\
LP 353-51 & M1.0 & 0 & 3772$\pm$115 & 4.79$\pm$0.18 & 1.59$\pm$0.32 & 4.0$\pm$2.2 \\
BD+02 348 & M1.0 & 1 & 3882$\pm$77 & 5.46$\pm$0.14 & $<$0.07 & 3.3$\pm$1.9 \\
HD 263175B & M1.0 & 1 & 3845$\pm$102 & 5.35$\pm$0.20 & $<$0.21 & 4.0$\pm$2.4 \\
BD+44 2051 & M1.0 & 2 & 3872$\pm$81 & 5.47$\pm$0.14 & $<$0.10 & 3.4$\pm$2.0 \\
2MASS J00393579-3816584 & M1.4 & 1 & 3535$\pm$116 & 4.55$\pm$0.17 & 2.54$\pm$0.32 & 9.0$\pm$1.9 \\
* tet Per B & M1.5 & 0 & 3584$\pm$96 & 4.76$\pm$0.16 & $<$0.25 & 2.8$\pm$1.9 \\
BD+18 3421 & M1.5 & 0 & 3856$\pm$108 & 5.38$\pm$0.18 & $<$0.07 & 5.5$\pm$1.9 \\
HD 31867B & M1.5 & 0 & 3589$\pm$182 & 4.71$\pm$0.30 & $<$0.36 & 4.5$\pm$2.8 \\
LP 298-53 & M1.5 & 0 & 3622$\pm$159 & 4.92$\pm$0.26 & 1.77$\pm$0.39 & 6.1$\pm$2.6 \\
NLTT 19115 & M1.5 & 0 & 3703$\pm$130 & 5.01$\pm$0.21 & $<$0.24 & 3.7$\pm$2.2 \\
Wolf 58 & M1.5 & 0 & 3648$\pm$153 & 4.69$\pm$0.22 & 1.82$\pm$0.35 & 6.8$\pm$2.4 \\
HD 154363B & M1.5 & 1 & 3771$\pm$108 & 5.13$\pm$0.21 & $<$0.12 & 3.6$\pm$2.1 \\
G 202-48 & M1.5 & 2 & 3852$\pm$94 & 5.46$\pm$0.14 & $<$0.15 & 5.1$\pm$2.3 \\
LP 493-31 & M1.5 & 2 & 3591$\pm$129 & 5.12$\pm$0.21 & $<$0.18 & 3.2$\pm$2.2 \\
BD+01 2447 & M2.0 & 0 & 3684$\pm$95 & 5.11$\pm$0.17 & $<$0.12 & 3.6$\pm$1.9 \\
BD+25 3173 & M2.0 & 0 & 3599$\pm$108 & 4.73$\pm$0.18 & $<$0.21 & 3.2$\pm$1.9 \\
BD+44 3567 & M2.0 & 0 & 3716$\pm$109 & 5.14$\pm$0.19 & $<$0.16 & 3.5$\pm$2.1 \\
HD 119850 & M2.0 & 0 & 3693$\pm$87 & 5.06$\pm$0.16 & $<$0.06 & 3.3$\pm$1.8 \\
HD 217987 & M2.0 & 0 & 3742$\pm$121 & 5.13$\pm$0.21 & $<$0.21 & 3.8$\pm$2.2 \\
V* AN Sex & M2.0 & 0 & 3550$\pm$124 & 4.67$\pm$0.20 & $<$0.32 & 3.3$\pm$2.1 \\
Ross 799 & M2.0 & 1 & 3476$\pm$156 & 4.61$\pm$0.26 & $<$0.36 & 4.0$\pm$2.4 \\
HD 349726 & M2.0 & 2 & 3876$\pm$80 & 5.46$\pm$0.14 & $<$0.09 & 3.9$\pm$2.1 \\
Ross 730 & M2.0 & 2 & 3875$\pm$80 & 5.47$\pm$0.14 & $<$0.08 & 3.1$\pm$1.9 \\
* 18 Pup B & M2.5 & 0 & 3585$\pm$156 & 4.87$\pm$0.28 & $<$0.32 & 5.2$\pm$2.6 \\
* tau Cyg C & M2.5 & 0 & 3696$\pm$144 & 5.16$\pm$0.26 & $<$0.35 & 4.7$\pm$2.8 \\
G 145-11 & M2.5 & 0 & 3686$\pm$166 & 4.98$\pm$0.31 & $<$0.38 & 4.7$\pm$3.1 \\
HD 285968 & M2.5 & 0 & 3477$\pm$128 & 4.64$\pm$0.20 & $<$0.36 & 3.5$\pm$2.2 \\
HD 38529B & M2.5 & 0 & 3404$\pm$178 & 4.47$\pm$0.27 & $<$0.39 & 3.9$\pm$2.4 \\
HD 50281B & M2.5 & 0 & 3633$\pm$109 & 5.00$\pm$0.18 & $<$0.26 & 3.5$\pm$2.1 \\
Ross 265 & M2.5 & 0 & 3512$\pm$112 & 4.67$\pm$0.21 & $<$0.41 & 8.3$\pm$2.0 \\
UCAC4 315-070111 & M2.5 & 0 & 3634$\pm$116 & 5.15$\pm$0.21 & 2.14$\pm$0.38 & 9.6$\pm$2.4 \\
G 155-29 & M2.5 & 1 & 3633$\pm$129 & 4.99$\pm$0.23 & $<$0.17 & 6.2$\pm$2.2 \\
LP 222-50 & M2.5 & 2 & 3776$\pm$115 & 5.35$\pm$0.21 & $<$0.22 & 3.8$\pm$2.3 \\
LP 636-19 & M2.9 & 1 & 3554$\pm$124 & 4.94$\pm$0.26 & $<$0.71 & 15.0$\pm$2.5 \\
G 121-42 & M3.0 & 0 & 3681$\pm$106 & 5.12$\pm$0.19 & $<$0.10 & 3.8$\pm$2.0 \\
GJ 752 & M3.0 & 0 & 3492$\pm$99 & 4.68$\pm$0.18 & $<$0.21 & 3.3$\pm$1.9 \\
GJ569 & M3.0 & 0 & 3651$\pm$129 & 4.99$\pm$0.22 & $<$0.99 & 5.1$\pm$2.5 \\
Ross 905 & M3.0 & 0 & 3569$\pm$107 & 4.91$\pm$0.18 & $<$0.14 & 3.2$\pm$1.9 \\
BD+68 946 & M3.0 & 1 & 3600$\pm$99 & 4.99$\pm$0.17 & $<$0.10 & 3.3$\pm$2.0 \\
G 80-21 & M3.0 & 1 & 3603$\pm$210 & 4.92$\pm$0.36 & 2.51$\pm$0.79 & 9.2$\pm$5.6 \\
HD 173739 & M3.0 & 1 & 3878$\pm$79 & 5.48$\pm$0.13 & $<$0.06 & 2.8$\pm$1.8 \\
UCAC4 595-047332 & M3.0 & 1 & 3675$\pm$149 & 5.26$\pm$0.26 & 1.67$\pm$0.55 & 7.0$\pm$3.3 \\
G 83-44 & M3.45 & 0 & 3494$\pm$116 & 4.77$\pm$0.21 & $<$0.19 & 3.5$\pm$2.1 \\
* ksi Peg B & M3.5 & 0 & 3749$\pm$139 & 5.23$\pm$0.26 & $<$0.31 & 5.0$\pm$2.9 \\
BD+05 1668 & M3.5 & 1 & 3746$\pm$109 & 5.32$\pm$0.19 & $<$0.06 & 3.5$\pm$1.9 \\
CD-44 11909 & M3.5 & 1 & 3433$\pm$107 & 4.94$\pm$0.22 & $<$0.15 & 7.1$\pm$2.0 \\
HD 127871B & M3.5 & 1 & 3713$\pm$154 & 5.24$\pm$0.26 & $<$0.24 & 4.6$\pm$2.7 \\
BD-15 6290 & M3.5 & 2 & 3467$\pm$101 & 5.06$\pm$0.18 & $<$0.16 & 2.7$\pm$1.9 \\
G 106-36 & M3.5 & 2 & 3680$\pm$152 & 5.28$\pm$0.29 & $<$0.31 & 6.7$\pm$2.9 \\
HD 173740 & M3.5 & 2 & 3882$\pm$78 & 5.49$\pm$0.13 & $<$0.06 & 4.5$\pm$2.1 \\
V* CW UMa & M3.5 & 2 & 3812$\pm$124 & 5.39$\pm$0.19 & 3.60$\pm$0.49 & 5.2$\pm$3.5 \\
Wolf 1062 & M3.5 & 2 & 3883$\pm$78 & 5.48$\pm$0.13 & $<$0.04 & 6.9$\pm$1.8 \\
Ross 104 & M4.0 & 0 & 3648$\pm$97 & 5.10$\pm$0.18 & $<$0.17 & 3.7$\pm$2.0 \\
BD+20 2465 & M4.0 & 1 & 3548$\pm$91 & 5.11$\pm$0.17 & 2.55$\pm$0.33 & 2.9$\pm$1.9 \\
G 96-10 & M4.0 & 1 & 3603$\pm$128 & 5.25$\pm$0.24 & $<$0.35 & 3.6$\pm$2.3 \\
HD 18143C & M4.0 & 1 & 3515$\pm$188 & 4.93$\pm$0.36 & $<$0.48 & 4.9$\pm$3.2 \\
LP 642-48 & M4.0 & 1 & 3614$\pm$137 & 5.21$\pm$0.24 & 3.44$\pm$0.36 & 14.2$\pm$2.3 \\
Ross 689 & M4.0 & 1 & 3570$\pm$125 & 5.05$\pm$0.21 & $<$0.18 & 3.6$\pm$2.2 \\
UCAC4 468-040412 & M4.0 & 1 & 3575$\pm$152 & 5.21$\pm$0.24 & $<$0.33 & 4.0$\pm$2.4 \\
V* WW PsA & M4.0 & 1 & 3578$\pm$131 & 4.89$\pm$0.26 & 2.39$\pm$0.53 & 16.1$\pm$2.6 \\
V* YZ CMi & M4.0 & 1 & 3606$\pm$111 & 5.36$\pm$0.18 & 3.81$\pm$0.29 & 3.3$\pm$2.0 \\
Wolf 358 & M4.0 & 1 & 3463$\pm$111 & 5.05$\pm$0.20 & $<$0.16 & 3.4$\pm$2.1 \\
Wolf 414 & M4.0 & 1 & 3493$\pm$106 & 4.95$\pm$0.22 & $<$0.13 & 8.0$\pm$2.0 \\
UCAC4 536-150368 & M4.0 & 2 & 3798$\pm$130 & 5.43$\pm$0.22 & 2.22$\pm$2.07 & 15.6$\pm$5.5 \\
2MASS J23263962+4521141 & M4.5 & 0 & 3766$\pm$149 & 4.95$\pm$0.32 & $<$0.62 & 9.5$\pm$4.1 \\
* rho01 Cnc B & M4.5 & 2 & 3391$\pm$108 & 4.89$\pm$0.21 & $<$0.14 & 2.8$\pm$1.9 \\
2MASS J04324938+2253082 & M4.6 & 1 & 3415$\pm$271 & 4.07$\pm$0.45 & $<$0.42 & 20.6$\pm$3.6 \\
V* TX PsA & M5.0 & 1 & 3720$\pm$187 & 5.04$\pm$0.36 & $<$0.33 & 27.2$\pm$2.7 \\
LP 214-42 & M5.0 & 2 & 3696$\pm$229 & 5.13$\pm$0.49 & $<$1.07 & 45.7$\pm$4.7 
\enddata
\end{deluxetable*}

\bibliography{oph}
\end{document}